\documentclass[twocolumn]{aastex62}

\newcommand{\ud}{\mathrm{d}}
\newcommand{\ie}{{\rm i.e.}}
\newcommand{\eg}{{\rm e.g.}}

\newcommand{\ed}{\varepsilon_{\mathrm{d}}}
\newcommand{\eb}{\varepsilon_{\mathrm{b}}}
\newcommand{\ee}{\varepsilon_{\mathrm{e}}}
\newcommand{\rd}{r_{\mathrm{d}}}

\newcommand{\rph}{r_{\mathrm{ph}}}
\newcommand{\Lum}{L_{0,52}}

\newcommand{\nr}{N_\mathrm{R}}
\newcommand{\pb}{p_{\mathrm{b}}}
\def\cf{{\rm cf.~\/}}

\graphicspath{{./}{figures/}}

\accepted{May 29, 2019}
\submitjournal{ApJ}

\usepackage{amsmath}	
\usepackage{amssymb}	
\usepackage{gensymb}

\shorttitle{Testing the DREAM}
\shortauthors{Ahlgren et al.}

\begin{document}

\title{Investigating subphotospheric dissipation in gamma-ray bursts using joint Fermi-Swift observations} 

\correspondingauthor{Bj\"orn Ahlgren}
\email{bjornah@kth.se}

\author[0000-0003-4000-8341]{Bj\"orn Ahlgren}
\affil{KTH Royal Institute of Technology, Department of Physics,\\ 
and the Oscar Klein Centre \\
AlbaNova, SE-106 91 Stockholm, Sweden}

\author{Josefin Larsson}
\affil{KTH Royal Institute of Technology, Department of Physics,\\ 
and the Oscar Klein Centre \\
AlbaNova, SE-106 91 Stockholm, Sweden}

\author{Vlasta Valan}
\affil{KTH Royal Institute of Technology, Department of Physics,\\ 
and the Oscar Klein Centre \\
AlbaNova, SE-106 91 Stockholm, Sweden}

\author{Daniel Mortlock}
\affil{Imperial College London, Department of Mathematics, \\
Statistics Section, London SW7 2AZ, UK}
\affil{Imperial College London, Astrophysics group, \\
Blackett Laboratory, Prince Consort Road, London SW7 2AZ, UK}
\affil{Stockholm University, Department of Astronomy,\\ 
and the Oscar Klein Centre \\
AlbaNova, SE-106 91 Stockholm, Sweden}

\author{Felix Ryde}
\affil{KTH Royal Institute of Technology, Department of Physics,\\ 
and the Oscar Klein Centre \\
AlbaNova, SE-106 91 Stockholm, Sweden}

\author{Asaf Pe'er}
\affil{University College Cork, Physics Department,
Cork, Ireland}



\begin{abstract}
The jet photosphere has been proposed as the origin for the gamma-ray burst (GRB) prompt emission. In many such models, characteristic features in the spectra appear below the energy range of the \textit{Fermi} GBM detectors, so joint fits with X-ray data are important in order to assess the photospheric scenario. Here we consider a particular photospheric model which assumes localized subphotospheric dissipation by internal shocks in a non-magnetized outflow. We investigate it using Bayesian inference and a sample of 8 GRBs with known redshifts which are observed simultaneously with \textit{Fermi} GBM and \textit{Swift} XRT. This provides us with an energy range of $0.3$~keV to $40$~MeV and much tighter parameter constraints. We analyze 32 spectra and find that 16 are well described by the model. We also find that the estimates of the bulk Lorentz factor, $\Gamma$, and the fireball luminosity, $\Lum$, decrease while the fraction of dissipated energy, $\ed$, increase in the joint fits compared to GBM only fits. These changes are caused by a small excess of counts in the XRT data, relative to the model predictions from fits to GBM only data. The fact that our limited implementation of the physical scenario yields 50\% accepted spectra is promising, and we discuss possible model revisions in the light of the new data. Specifically, we argue that the inclusion of significant magnetization, as well as removing the assumption of internal shocks, will provide better fits at low energies.
\end{abstract}



\section{Introduction} \label{sec:intro}
The prompt phase of gamma-ray bursts (GRBs) is characterized by strongly variable gamma-ray emission that typically lasts less than a minute. While almost all models agree that this emission originates from internal processes in a relativistic jet, the mechanism producing the emission is not understood. 
GRB spectra are often described using empirical models, particularly a smoothly broken power law known as the Band function \citep{1993ApJ...413..281B}. This function has often been interpreted in terms of synchrotron radiation, see \eg ~\cite{1996ApJ...466..768T,Briggs:1999tg,2009Sci...323.1688A,2016ApJ...816...72Z}. However, it was argued by \cite{Preeceetal_1998A_ApJ} that fits with the Band function show that a large fraction of observed spectra are harder than can be accounted for by synchrotron radiation. 
This has been one of the reasons for considering other emission mechanisms. 
There are also some GRBs observed with very hard spectra, which can be well described by blackbody or multi-color blackbody models, \citep{2004ApJ...614..827R,2011MNRAS.415.3693R,Ghirlandaetal_2013A_MNRAS,2015ApJ...800L..34L}. Although pure blackbody emission is clearly too hard to describe most observed spectra, there are several possible broadening mechanisms which soften the low-energy slope of photospheric emission, including geometric effects \citep{2008ApJ...682..463P,Lundmanetal_2013A_MNRAS}, and subphotospheric dissipation \citep{2005ApJ...628..847R, 2006ApJ...642..995P,2006A&A...457..763G,2010MNRAS.407.1033B,2011ApJ...738...77V,2015ApJ...802..132C}. 
Photospheric emission also provides a viable explanation for the Yonetoku relation \citep{2004ApJ...609..935Y,2010PASJ...62.1495Y}, as shown by \cite{2018arXiv180600590I}. See also \cite{2018ApJ...869..103P} for further discussion on photospheric emission as an origin of the Yonetoku relation, as well as for the related Amati \citep{2002A&A...390...81A} and Golenetskii \citep{1983Natur.306..451G} relations.

To evaluate different physical models for the prompt emission it is important to directly fit the models to data. Indeed, considerable work in recent years have shown that inferences about emission processes based on fits with the Band function, including the aforementioned hardness-problem, can be misleading \citep{2014ApJ...784...17B,2017ifs..confE..74B,2018arXiv181006965B,Ahlgren:2019krn}. Examples of previous physical models fitted to data include a physical synchrotron model (fit to GRBs observed by the BATSE instrument; \citealt{2000ApJ...543..722L}), the ICMART model (fit to GRB~080916C; \citealt{2011ApJ...726...90Z}), and the external shock model (fit to GRB~141028A; \citealt{2016ApJ...822...63B}). Unfortunately, fitting these models to data is generally time consuming, making broad usage difficult. However, there have recently been new development, with \cite{2018arXiv181006965B} showing successful fits to a sample of 19 GRBs using a physical synchrotron model.

Another recent example is provided in \cite{Ahlgren:2019krn} (A19 from now on), where we tested a specific photospheric model for localized subphotospheric dissipation by internal shocks with no magnetic fields (DREAM, introduced in \citealt{2015MNRAS.454L..31A}; A15 from now on). We analyzed time-resolved spectra of 36 GRBs observed by the Gamma-ray Burst Monitor (GBM; \citealt{2009ApJ...702..791M}) on board the \textit{Fermi Gamma-ray Space Telescope} and found that $\sim ~30\%$ of the spectra could be well described by the model, with 10 GRBs having at least half of their spectra accepted. The model consistently failed to describe the GRBs with the highest luminosities, which may be a result of the specific dissipation scenario considered. The level of dissipation was found to be around $\sim ~5\%$, consistent with the internal shock scenario, and the luminosity and Lorentz factor of the model were found to be correlated. The latter has independently been reported in other studies \citep{2012ApJ...751...49L,2018A&A...609A.112G}. In our study there was no correlation between the fitted Band function $\alpha$ and $\beta$ parameters and any properties of the fits using the DREAM model. The current implementation of the physical scenario is still being improved upon, and the large number of well-described spectra motivates further exploration of the model scenario.

While \textit{Fermi} observes GRB prompt emission over a wider energy range than any other GRB mission, it is limited by its lower energy bound of $8$~keV. Physical models (including DREAM) often predict distinct spectral features below this energy and observations of the prompt emission down to soft X-rays therefore have the potential to be very constraining. 
For instance, in the case of DREAM, the curvature at low energies is a result of incomplete Comptonization of the seed photon blackbody, which may be located at energies as low as $\sim$1~keV in the observer frame. 
The X-ray Telescope (XRT; \citealt{2005SSRv..120..165B}) on board the \textit{Neil Gehrels Swift Observatory} observes in the $0.3-10$~keV energy band, but the observations are limited by the fact that they typically start $\sim 100$~s after a trigger from the Burst Alert Telescope (BAT). For most GRBs the prompt gamma-ray emission has already ended by this time. 

Studies of early XRT observations have shown that the emission is due to a combination of late prompt emission and early afterglow emission from the interaction between the jet and circumstellar medium \citep{2006ApJ...647.1213O,2007RSPTA.365.1213B}. The former is manifested by flares in the light curve, which have properties similar to the prompt gamma-ray emission \citep{2010MNRAS.406.2113C}. Given these considerations, we note that it is possible to study the prompt emission from soft X-rays to gamma rays in GRBs for which the \textit{Swift}/XRT light curve is dominated by flares and the prompt gamma-ray emission has a sufficiently long duration.

In previous joint analyses of XRT and GBM prompt emission, \cite{2017ApJ...846..137O,2018A&A...616A.138O} found that many spectra exhibit a second spectral break at around a few keV. They performed fits using a doubly smoothly broken power-law, and interpret their results in terms of synchrotron radiation. Additionally, \cite{2017A&A...598A..23N} performed a time resolved joint analysis of data from several instruments, including GBM and XRT, of GRB~151027A, where they detect the presence of a thermal component at low energies. 
These different results are interesting also from the perspective of photospheric models, as these often have a curvature at energies around a few keV.

In this work we present further investigation of the model presented in A19 by performing joint analysis to data from \textit{Fermi}/GBM and \textit{Swift}/XRT using Bayesian inference. The only difference to the model presented in A19 is a small expansion of the parameter space. We analyze a sample of 8 GRBs which have overlapping GBM and XRT observations and known redshifts. The goal is to use the large energy window offered by the joint observations to provide new constraints on the physical scenario and parameter space of our model. We also want to assess the impact of the XRT data in prompt emission analysis.

In Section~\ref{sec:model} we describe the physical scenario and how it is implemented as a numerical model. We describe the data sample and analysis in 
Sections~\ref{sec:datasample} and \ref{sec:analysis}, followed by a presentation of the results in Section~\ref{sec:results} and discussion in Section~\ref{sec:discussion}. In Section~\ref{sec:summary} we summarize our findings.
Throughout the paper we assume standard $\Lambda$-CDM cosmology using the constants $H_0 = 67.3$, $\Omega_{\lambda} = 0.685$, and $\Omega_\mathrm{M} = 0.315$ \citep{2014A&A...571A..16P}.

\section{Model}\label{sec:model}
We study a photospheric emission model, in which localized subphotospheric dissipation occurs due to internal shocks. Moreover, we assume no magnetization and ignore off-axis emission and other geometrical effects.
The model used in this work is identical to that of A19, apart from a small expansion of the parameter space. For completeness, we here briefly describe the physical scenario and its numerical implementation. For a more detailed description of the model, including validation tests, see A19 and the references given below.

\subsection{Physical scenario}\label{sec:physscenario}
The physical scenario we consider is a hot fireball (for an overview, see \eg~\citealt{2015AdAst2015E..22P}), with localized subphotospheric dissipation at a moderate optical depth. 
We assume that a central engine emits a hot plasma of baryons, electrons and photons, at a luminosity $L_0 = \Lum 10^{52}$ erg s$^{-1}$. The outflow accelerates due to the thermal pressure from the photons until it obtains its coasting bulk Lorentz factor, $\Gamma$. We assume a dissipation radius, $\rd = r_{\mathrm{ph}}/\tau = L \sigma_{\mathrm{T}} / (4\pi \tau \Gamma^3  c^3 m_{\mathrm{p}})$, where $\rph$ is the photospheric radius and $\tau$ is the optical depth (see \eg ~\citealt{2006ApJ...642..995P}). The internal shock assumption furthermore implies that $\rd = \Gamma^2 r_0$, where $r_0$ is the nozzle radius. This relation couples the photon temperature at the dissipation site to $\Gamma$, $\Lum$ and $\tau$. At $\rd$, a fraction $\ed\ee$ of the bulk kinetic energy of the outflow begins to dissipate to the electrons, and a fraction $\ed\eb$ to the magnetic fields. The dissipation is assumed to continue until $2\rd$. In the particular scenario considered here, we assume that $\eb = 10^{-6}$ and $\ee = 0.9$, \ie ~that magnetic fields are negligible and that almost all the energy is dissipated to the electrons.We also assume that the heated electrons are Maxwellian distributed. The remaining $\sim 0.1 \ed$ energy is considered not dissipated. The parameterization with $\ed$, $\ee$ and $\eb$ is for practical reasons only and that the total efficiency in this case is given by is $\ed\ee$. Further, we assume that $\tau = 35$, which means that we test a scenario where the dissipation occurs moderately deep below the photosphere. Tests have shown that this is often a good approximation since many bursts are largely insensitive to this parameter (see A19 for further discussion).

Since we assume a scenario where the dissipation occurs below the photosphere, the heated electrons will interact with the photon field in a non-equilibrium situation. In our model we consider Compton and inverse Compton scattering, pair production as well as pair annihilation. In principle we also account for synchrotron and synchrotron self-absorption. However, these effects are very small in the case of negligible magnetization.

For localized dissipation in outflows with negligible magnetization, hadronic collisions \citep{2010MNRAS.407.1033B} and internal shocks \citep{1997ApJ...490...92K,1998MNRAS.296..275D,2005ApJ...628..847R} have been suggested as possible dissipation mechanisms. The scenario we consider here is based on the internal shock scenario \citep{2015AdAst2015E..22P}.

To simulate the physical scenario we use the kinetic code by \cite{2005ApJ...628..857P}, which treats all processes described above. Our treatment does not include any spatial effects, such as geometric broadening \citep{2008ApJ...682..463P,Lundmanetal_2013A_MNRAS}, or jet hydrodynamics \citep{2009ApJ...700L..47L}. We also assume that the photosphere is sharp, as opposed to a fuzzy photosphere \citep{2011ApJ...737...68B,2013ApJ...767..139B}. 
While we restrict ourselves to a specific dissipation scenario, alternative scenarios are also possible, see \eg ~\cite{2011ApJ...738...77V}.

\subsection{Table model}\label{sec:tablemodel}
In order to perform fits with the model we construct an {\scriptsize XSPEC} compatible table model \citep{1999ascl.soft10005A}, which consists of a grid of spectra simulated for different parameter values. Model predictions for parameters between the grid points are obtained by linear interpolation during the fitting.
The simulations are costly to perform, which means that we are not able to explore all available parameter space. In A19 we presented a model in 3 dimensions with 891 grid points, consisting of the level of dissipation, $\ed$, the luminosity, $\Lum$, and the bulk Lorentz factor, $\Gamma$.

For this study we have expanded the model with additional grid points in $\Gamma$, extending the range down to $50$ from $100$ and from $500$ up to $1000$. We added these grid points to obtain a better coverage of values of $\Gamma$ inferred from observations \citep{2018A&A...609A.112G}. We have also added one additional grid point in $\Lum$, at $1000$. This grid point has been added to account for the possibility of very luminous bursts with narrow jets. There are no differences in the underlying code used to construct the grid. The model is now spanned by
\begin{align*}
\Gamma =& ~50,100,150,200,250,300,350, \\ &~400,450,500,600,700,800,900,1000\\
\Lum =& ~0.1, 0.5, 1, 5, 10, 50, 100, 200, 300, 1000 \\
\varepsilon_{\mathrm{d}} =& ~0.01,0.025,0.05,0.75,0.1,0.15, \\  & ~0.2,0.25, 0.3,0.35, 0.4.
\end{align*}
As noted in A19, high values of $\ed$ would yield a re-acceleration of the outflow, which is not accounted for in our code. Thus, we limit $\ed$ to values where the effects of such a re-acceleration are expected to be small.
In A19 we demonstrated that the finite resolution of the grid introduces systematic uncertainties of $5-8\%$ in the best-fit parameter values, which is smaller than the typical statistical uncertainties.  
When creating a table model we also obtain one parameter for the redshift and one for the normalization. The latter is set by the redshift as ${d_\mathrm{L}^2(z=1)}/{d_\mathrm{L}^2(z_\mathrm{obs})}$, where $d_\mathrm{L}(z)$ is the luminosity distance. Both of these parameters are kept fixed in the fits.

In line with previous work, we denote this version of the table model DREAM1.3, using the naming convention introduced in A19.\footnote{This naming convention was introduced because of the plan to make the model publicly available. It is then convenient to be able to distinguish between different versions of the model used in different articles.}
Throughout this work, `model' refers to DREAM1.3, unless otherwise stated. The underlying physical model scenario is referred to explicitly as `physical scenario', to avoid confusion. The physical scenario we are testing is subphotospheric dissipation with localized dissipation at a moderate optical depth in a jet with negligible magnetization. 

\section{Observations} \label{sec:datasample}
\subsection{Sample selection} \label{sec:sampleselection}
\begin{figure*}
    \centering
    \includegraphics[scale=0.48]{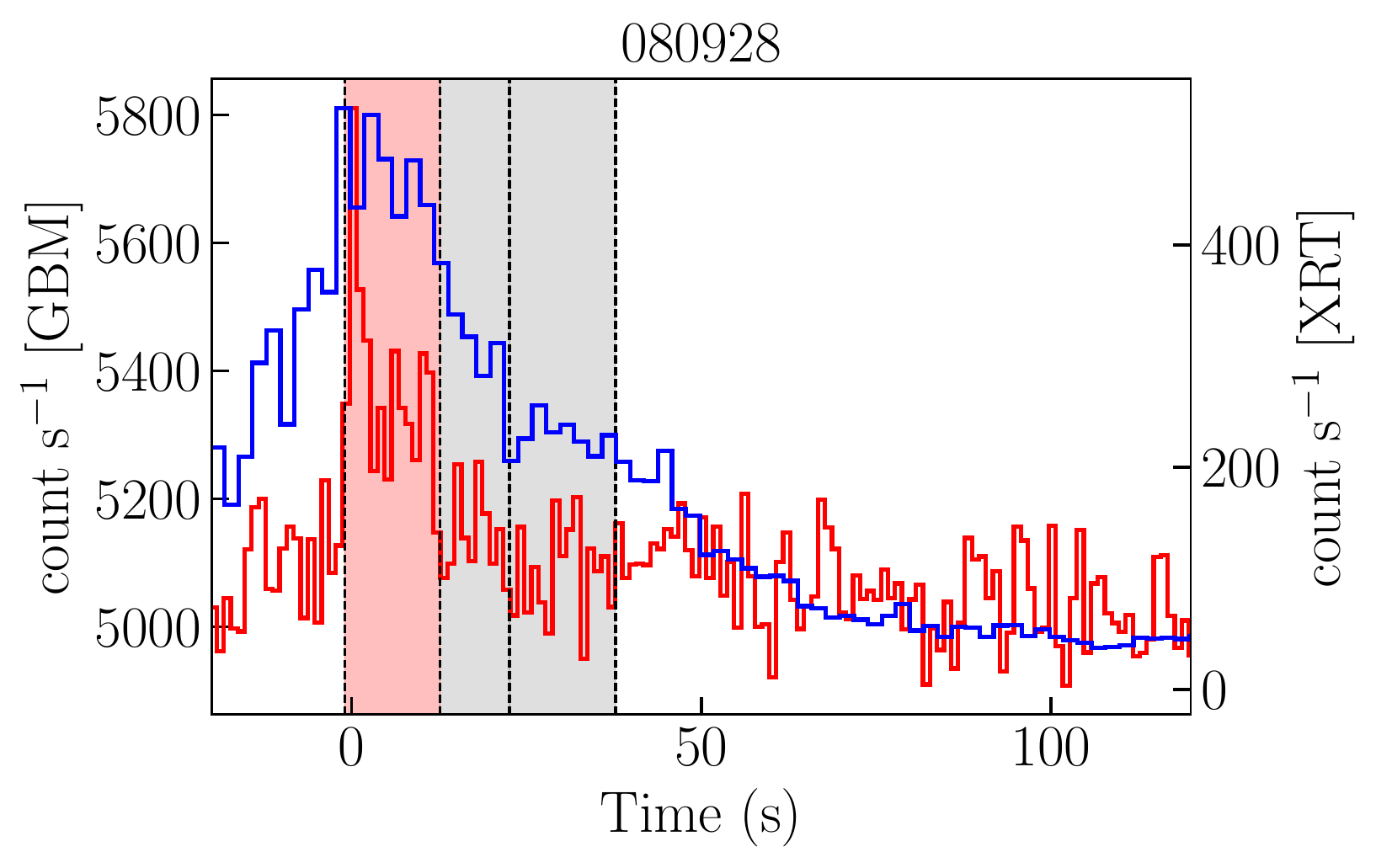}
    \includegraphics[scale=0.48]{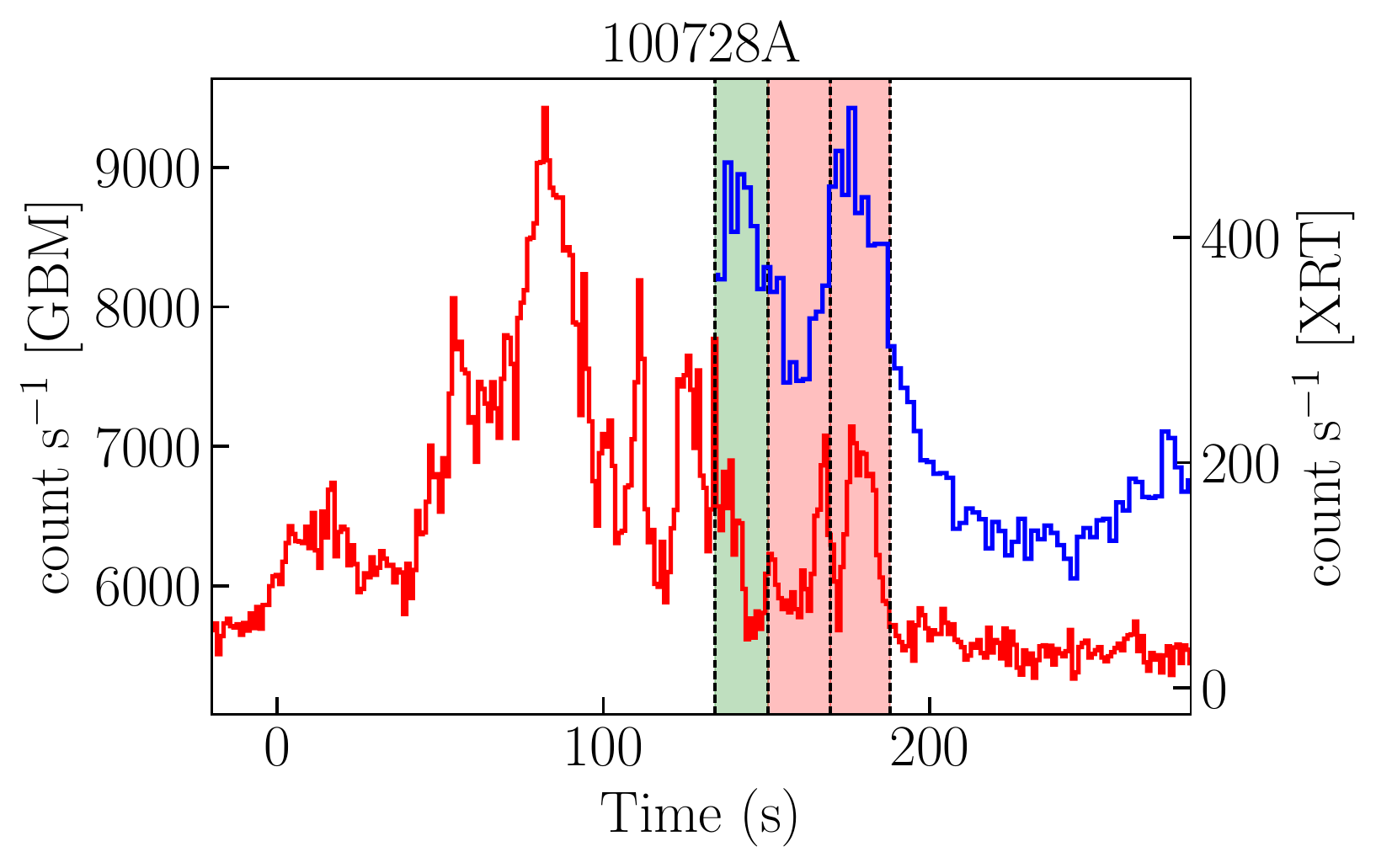}
    \includegraphics[scale=0.48]{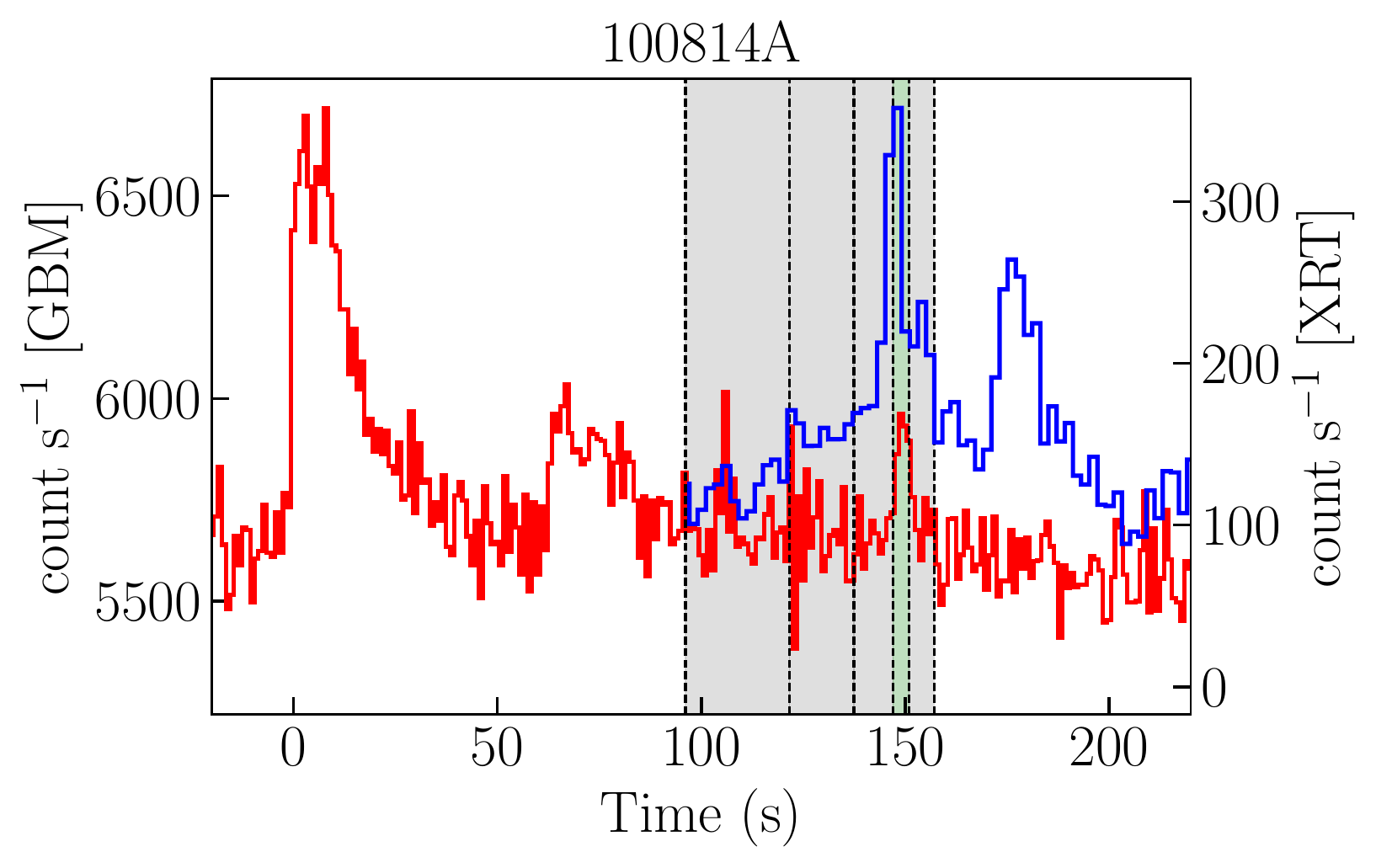}
    \includegraphics[scale=0.48]{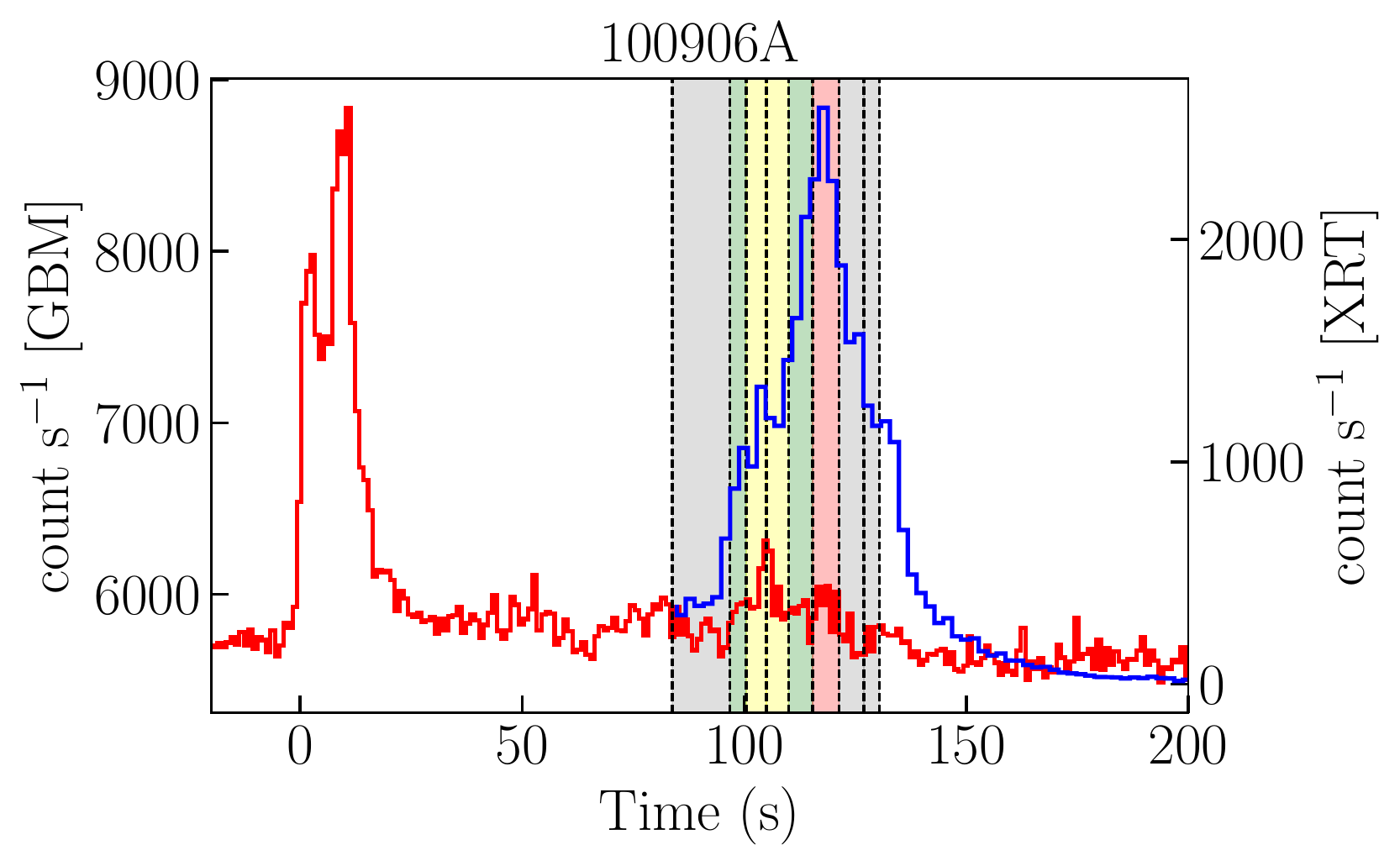}
    \includegraphics[scale=0.48]{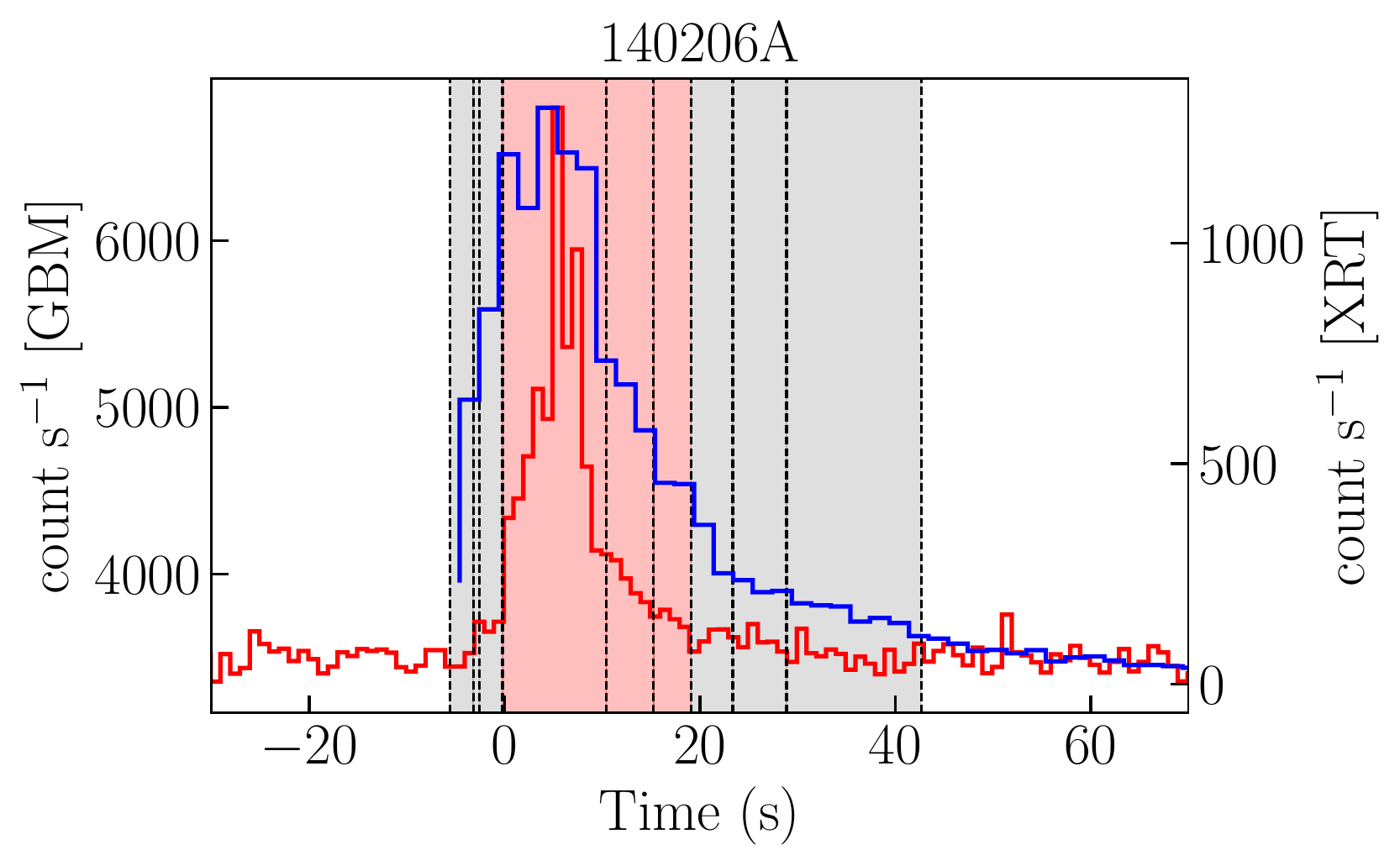}
    \includegraphics[scale=0.48]{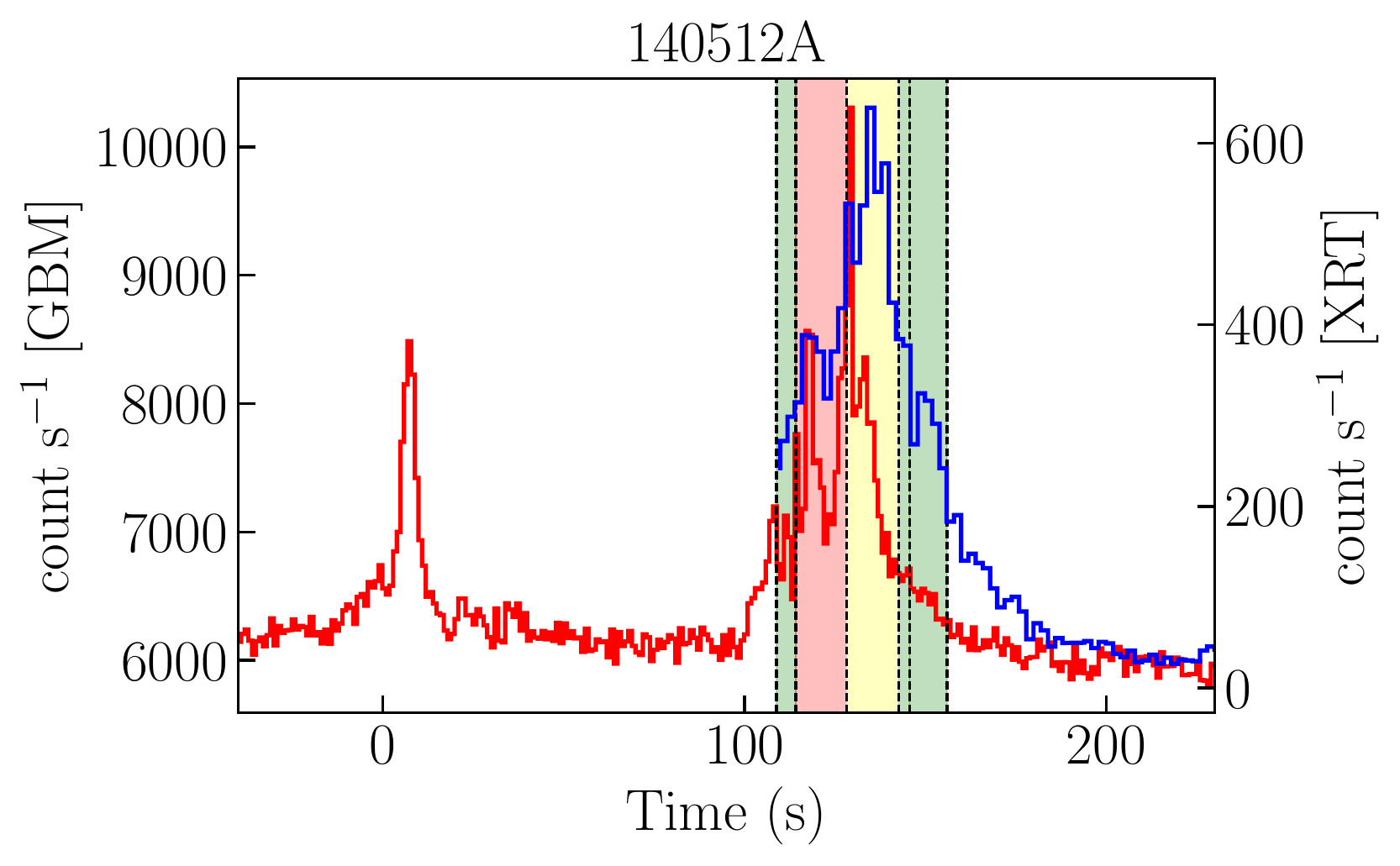}
    \includegraphics[scale=0.48]{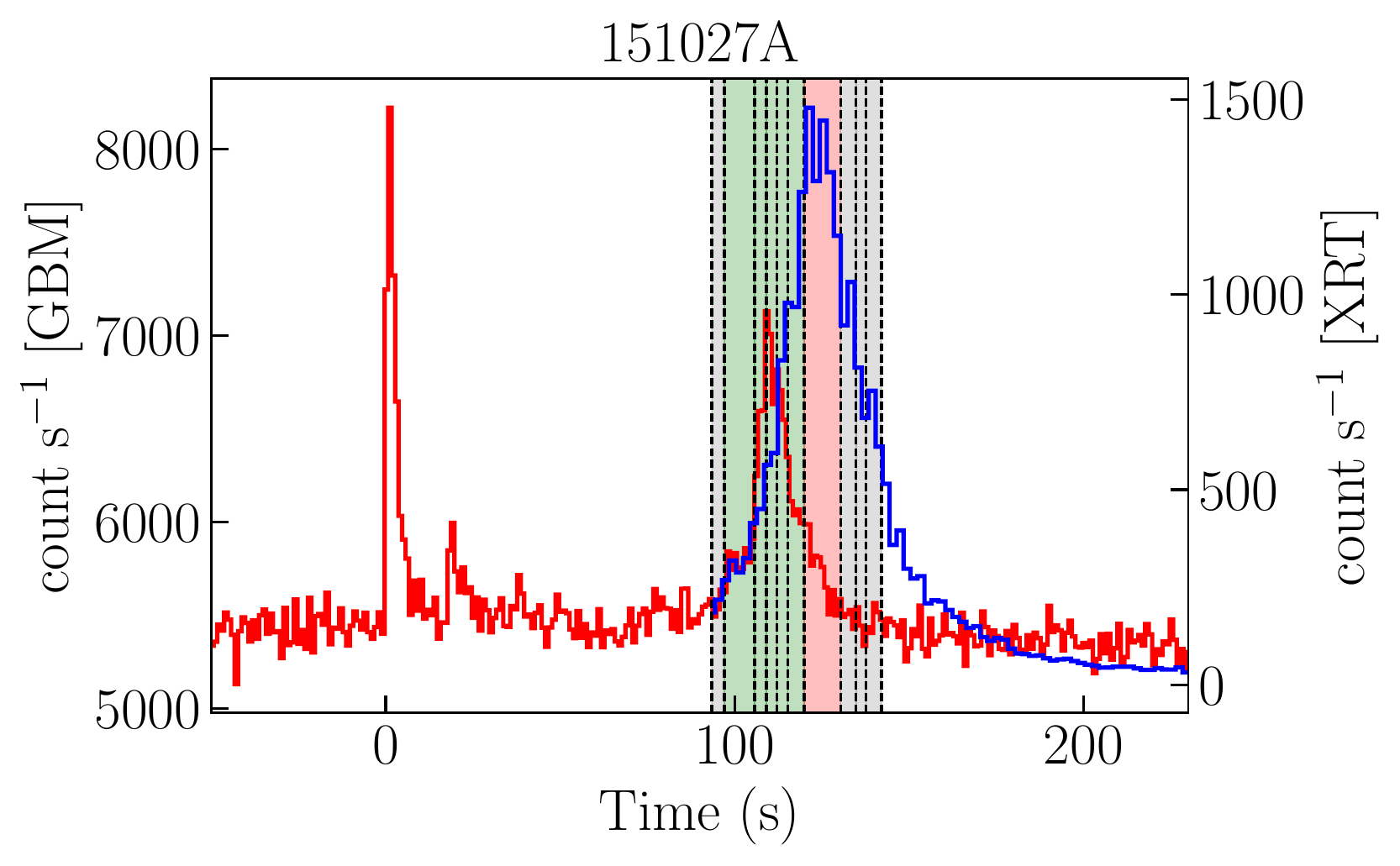}
    \includegraphics[scale=0.48]{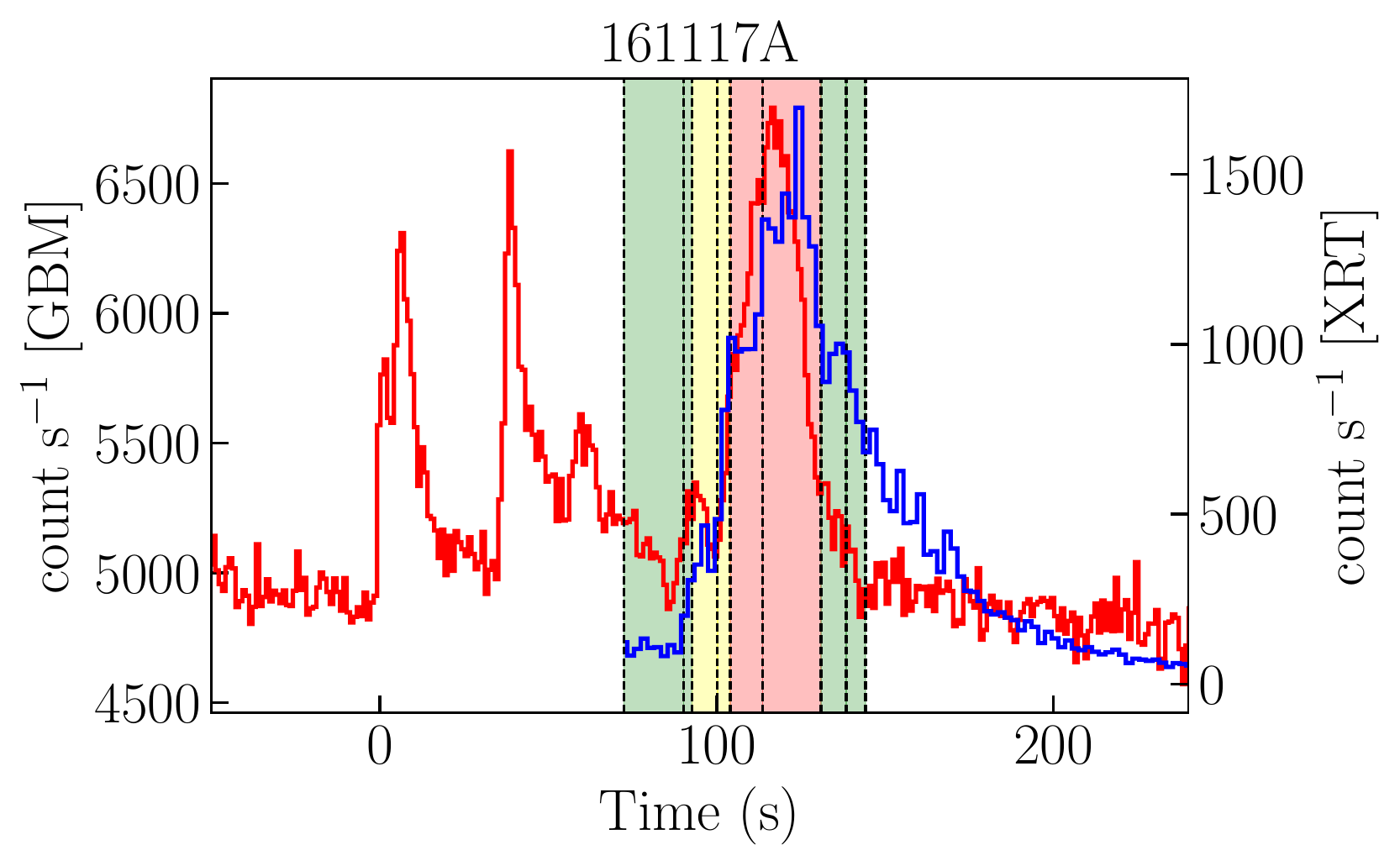}
    \caption{Count rate light curves for all GRBs in our sample. GBM and XRT data are marked in red and blue, respectively. The time refers to the GBM trigger. The count rate for the GBM is the summed count rate for all NaI and BGO detectors used in the analysis. The GBM and XRT light curves are binned in 1 and 2 second bins, respectively. The start and end time for all time bins identified by the Bayesian blocks algorithm in the overlapping region are marked by dashed black lines. Gray shading indicates that the bin has not been analyzed due to low SNR in the GBM (see Section~\ref{sec:analysis}). Green shading indicates that the fit passed the PPC (see Section~\ref{sec:ppp}). Similarly, red shading indicates that the fit did not pass the PPC. Yellow shading indicates that the fit passed the PPC but is considered rejected on other grounds (see Sections~\ref{sec:XRTimpact} and~\ref{sec:discuss:XRTimplications:AnalysisUncertainties}). Note that the XRT light curves are background subtracted, whereas the GBM light curves include background.}
    \label{fig:countRateLCs}
\end{figure*}

We examine all GRBs with a known redshift which have overlapping observations in the \textit{Swift} XRT and \textit{Fermi} GBM 
detectors, up until 2018-11-01. As mentioned in Section~\ref{sec:tablemodel}, the redshift is needed in order to obtain the luminosity distance for the model.
We also require the overlapping interval to be at least 5 seconds long and that it is possible to bin the XRT data into at least two time bins using the method outlined in Section~\ref{sec:analysis} below. Finally, we require that at least one time bin has a signal-to-noise ratio (SNR) in one GBM NaI detector of at least $3$, as described in Section~\ref{sec:analysis} (the SNR in the XRT is always higher than this). Note that we do not use data from the BAT, also on board the \textit{Swift} satellite, due to its narrower energy range ($15 - 150$~keV) compared to GBM.
Contrary to the burst sample in A19 we do not perform a fluence cut in this sample selection.
These criteria result in a relatively small sample of 8 GRBs, the properties of which are summarized in Table~\ref{tab:DataSample}. In Fig.~\ref{fig:countRateLCs}, we show the count rate light curves of all bursts, including both GBM and XRT data. 

XRT observations typically start $\sim 100$ s after the GBM trigger (corresponding to the time it takes for \textit{Swift} to slew following a BAT trigger). For most GRBs in the sample the GBM and XRT data overlap for $40-50$~s at the end of the prompt emission. However, there are two exceptions: GRB~080928 and GRB~140206A. These bursts have XRT data from the start of the GBM trigger. This is because BAT triggered on precursors at $-204$~s and $-56$~s, respectively, relative to the GBM trigger (not included in Fig.~\ref{fig:countRateLCs}). Fig.~\ref{fig:countRateLCs} also shows that all the XRT light curves contain flares, as expected if the emission belongs to the prompt phase. In most cases the GBM and XRT light curves are clearly correlated. However, we note that this correlation is less prominent in GRB~100814A and GRB~100906A. This could be due to that the GBM data are particularly weak in these intervals. Additionally, in GRB~151027A, there is a clear delay in the XRT with respect to the GBM. 

{\sloppy Comparing with the sample in A19, six bursts (GRB~100728A, GRB~100814A, GRB~100906A, GRB~140206A, GRB~140512A, and GRB~151027A) are contained in both samples. In A19 we found that only two of these bursts, GRB~140512A and GRB~151027A, had more than half of their analyzed time bins well described by our model. We discuss this further in section~\ref{sec:discussion:paper2}.}

\begin{table*}[]
    \centering
    \begin{tabular}{ccccccccc}
        \hline
        GRB & Redshift & GBM Detectors & $T_{90,\mathrm{GBM}}$ & $T_{0,\mathrm{XRT}}$\tablenote{With respect to the GBM trigger.} & Overlap & $N_{\mathrm{H,gal}}$ & $N_{\mathrm{H,intr}}$ & PC data interval \\  & & & (s) & (s) & (s) & ($10^{22}$ cm$^{-2}$) & ($10^{22}$ cm$^{-2}$) & (s) \\ \hline 
        080928  & 1.692 & NaI3, NaI6, NaI7, BGO0     & 14.3  &  - 27.2    & $37$ &  $0.072$ & $0.47_{-0.16}^{+0.16}$ & $4262 - 12612$ \\
        100728A & 1.567 & NaI0, NaI1, NaI2, BGO0     & 165.4 & 134.2      & $44$ &  $0.165$ & $3.16_{-0.42}^{+0.44}$ & $1333 - 7548$\\ 
        100814A & 1.44  & NaI7, NaI8, BGO1         & 150.5 & 96.1         & $55$ &  $0.018$ & $0.00_{}^{+0.09}$ & $5923 - 34841$ \\  
        100906A & 1.727 & NaI8, NaI11, BGO1        & 110.6 & 83.8         & $27$ &  $0.353$ & $1.24_{-0.64}^{+0.63}$ & $10592 - 99797$\\  
        140206A & 2.73  & NaI10, NaI11, BGO1       & 27.3  & -5.6         & $34$ &  $0.070$ & $1.08_{-0.44}^{+0.51}$ & $126574 - 956006$ \\  
        140512A & 0.725 & NaI0, NaI1, NaI3, BGO0     & 148.0 & 108.9      & $41$ &  $0.147$ & $0.31_{-0.05}^{+0.06}$ & $4581 - 12859$\\  
        151027A & 0.81  & NaI0, NaI3, BGO0             & 123.4 & 93.4     & $18$ &  $0.037$ & $0.58_{-0.13}^{+0.13}$ & $11633 - 63739$\\  
        161117A & 1.549 & NaI1, NaI2, NaI10, BGO0    & 122.2 & 72.5       & $53$ &  $0.043$ & $0.99_{-0.34}^{+0.38}$ & $46352 - 69247$\\  
         \hline
    \end{tabular}
    \caption{Summary of data sample, including which GBM detectors were used for the different bursts and for how long we have overlapping data of GBM and XRT. The overlap indicates the overlap between the start of the XRT data and the \textit{Fermi} $T_{90}$. Redshifts were acquired from \url{http://www.mpe.mpg.de/~jcg/grbgen.html.} We also present the values of the column densities, $N_{\mathrm{H,gal}}$ and $N_{\mathrm{H,intr}}$. The PC data interval indicates the time interval after the BAT trigger used when determining $N_{\mathrm{H,intr}}$.} 
    \label{tab:DataSample}
\end{table*}

\subsection{Data reduction} \label{sec:datareduction}

\subsubsection{Fermi data} \label{sec:GBMreduction}
\begin{figure}
    \centering
    \includegraphics[scale=0.35]{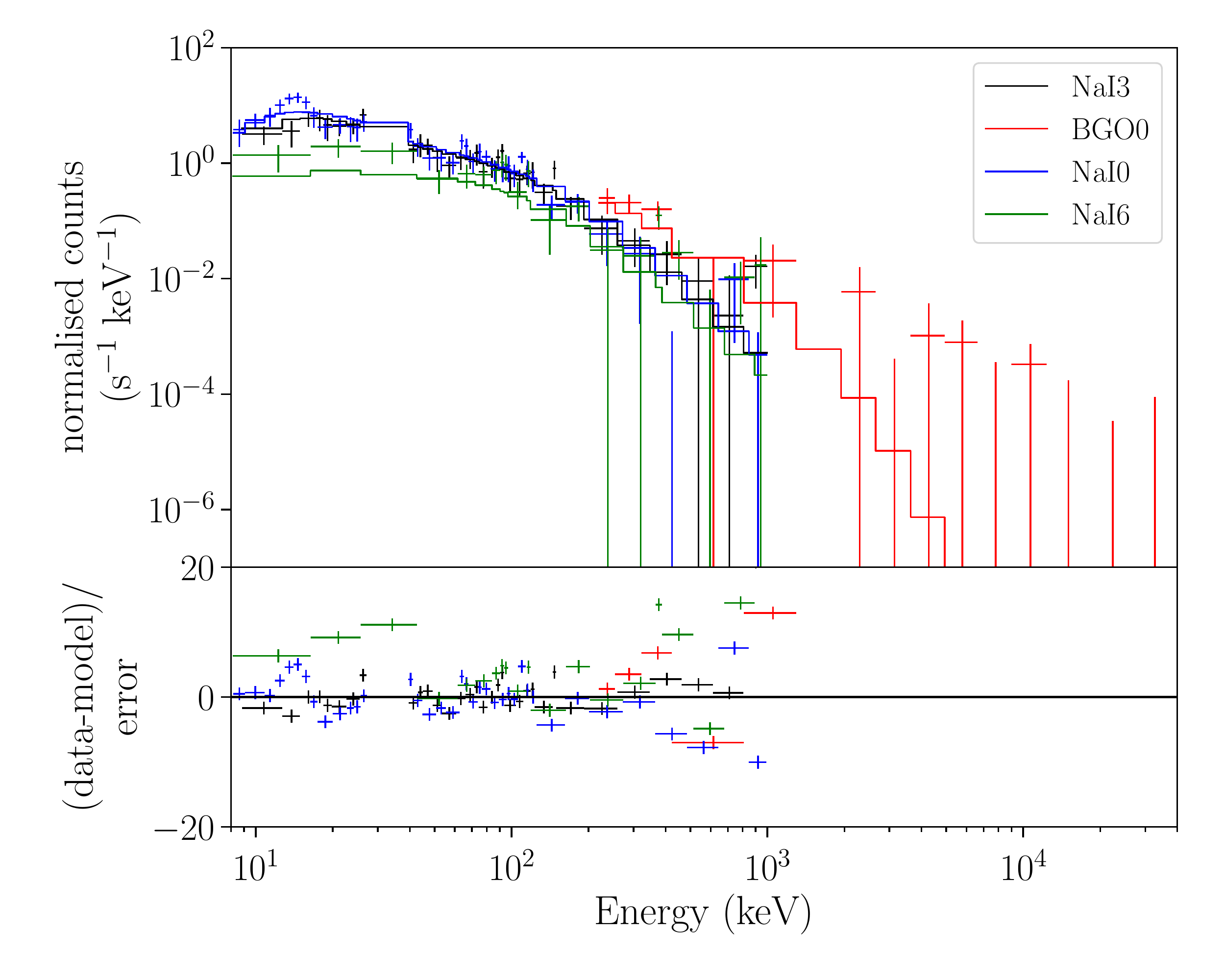}
    \caption{The counts spectrum of GRB~151027A in the time interval $115.2 - 120.0$~s, fitted with a cutoff power law model. The detectors corresponding to the different data sets are given by the legend. There are large and systematic discrepancies between the best fit model and the data in the NaI6 detector. This is caused by some unknown issue with the detector response. This detector is therefore excluded from the analysis.}
    \label{fig:badDetector}
\end{figure}

In the \textit{Fermi} analysis we use data observed with the GBM. Specifically we use the Time-Tagged Event (TTE) data from both the NaI and BGO detectors. We include up to three NaI detectors with an angle of incidence less than $60 \degree$ in the analysis, as well as the BGO detector with the lowest angle of incidence \citep{2016yCat..22230028B}. However, there is one exception where we have excluded an additional NaI from the analysis. 
In GRB~151027A we use only the NaI0 ($19 \degree$) and NaI3 ($37 \degree$) detectors, ignoring the n6 ($25 \degree$) detector, which has a significantly different spectrum in several bins (example in Fig.~\ref{fig:badDetector}). This may be due to blockage of the detector by a part of the satellite, or some other issue with the detector response \citep{2012ApJS..199...19G}.

The background is determined from a polynomial fit to the light curve, which gives a model for the background with Gaussian errors. We account for temporal evolution of the background spectrum and the changing position of the spacecraft by using rsp2 response files when available.

\subsubsection{XRT data} \label{sec:XRTreduction}
For the spectral analysis of XRT data we use observations taken in the Windowed Timing (WT) mode, which is used when the count rate is high. We also download late time Photon Counting (PC) data for each GRB, in order to determine the intrinsic column density ($N_\mathrm{H,intr}$) for the absorption at low energies (see Section~\ref{sec:absorption}).

The WT data are downloaded as locally-reprocessed data from the UK \textit{Swift} Science Data Centre XRT GRB repository\footnote{\url{http://www.swift.ac.uk/xrt_live_cat/}}. We create time bins of the spectrum locally, as described in Section~\ref{sec:analysis}. These time bins are then used to specify the limits on time-sliced spectra, which we create and download from the online repository. All spectra are grouped such that each energy bin contains at least one count. This the recommended approach when fitting with the cstat fit statistic in {\scriptsize XSPEC}\footnote{see the {\scriptsize XSPEC} manual appendix \url{https://heasarc.gsfc.nasa.gov/xanadu/xspec/manual/XSappendixStatistics.html}}. The background is supplied online when downloading the data. The background spectrum is constructed by sampling in an area around the burst position and it is assumed to be Poisson distributed.

All XRT spectra were checked for known calibration issues following \cite{2018MNRAS.474.2401V}. This includes redistribution issues, which may cause a bump below 1 keV and/or a turn-up below $\sim$0.6~keV. Pile-up is automatically dealt with by the online tool. We find redistribution issues in GRB~100728A and GRB~151027A. We accommodate these by ignoring channels below $0.6$~keV in these bursts. 

\section{Data Analysis} \label{sec:analysis}
We perform a time resolved spectral analysis using time bins defined by a Bayesian blocks binning \citep{2013arXiv1304.2818S} of the XRT data. We use {\scriptsize battblocks}\footnote{Heasoft version 6.17} with default settings to bin the XRT light curve. The GBM data are binned in matching time bins. We use HEASARC's online tool xTime\footnote{\url{https://heasarc.gsfc.nasa.gov/cgi-bin/Tools/xTime/xTime.pl}} to convert between \textit{Swift} mission time and \textit{Fermi} mission time. 

We calculate the SNR of each spectrum as described in \cite{2018ApJS..236...17V}. For the GBM data we use the Poisson-Gaussian significance, whereas we for the XRT data use the Poisson-Poisson significance. The XRT data consistently have a high SNR, but as in A19 we choose to apply an SNR cut to the GBM data in order to only analyze spectra which contain a significant signal in GBM. Thus, we only analyze spectra with SNR$> 3$ in the brightest NaI detector. This leaves us with 32 out of 51 time-resolved spectra to analyze.

\subsection{Fitting}\label{sec:fitting}
We set up the analysis as a Bayesian inference procedure. Bayes' theorem states that the posterior probability is
\begin{align*}
    \Pr (\theta | y) = \frac{\Pr (\theta) \Pr (y | \theta)}{\int \Pr (\theta) \Pr (y | \theta) \ud \theta} \propto \Pr (\theta) \Pr (y | \theta),
\end{align*}
where $\theta$ are our model parameters and $y$ the observed data. $\Pr (\theta)$ is the prior, $\Pr (y|\theta)$ the likelihood, and the denominator the marginalized likelihood (also referred to as the evidence). We use PyMultiNest \citep{2014A&A...564A.125B}, a python implementation of MultiNest \citep{2008MNRAS.384..449F,2009MNRAS.398.1601F,2013arXiv1306.2144F}, to sample from the model posterior using 600 live points. We have chosen this particular number by testing the analysis with different number of live points to ensure stability.

For the \textit{Fermi} data, we consider the energy range 8-1000 keV and 200 keV - 40 MeV for the NaI and BGO detectors, respectively. We also ignore the interval 30-40 keV in the NaI detectors, because of the iodine $K$-edge \citep{2009ExA....24...47B}. For the XRT data we consider the nominal energy range 0.3-10 keV. 
However, we lower the high-energy limit in cases when the signal stops below 10 keV. Additionally, the low-energy limit is modified in the presence of calibration issues, as described in Section~\ref{sec:XRTreduction} (which affects GRB~100728A and GRB~151027A).

For the \textit{Fermi} data we use a likelihood for a Poisson distributed signal with Gaussian background. In {\scriptsize XSPEC} the corresponding statistic is known as pgstat (for a description of pgstat, see \eg ~\citealt{Burgess:2018get}). For the XRT data we adopt the Cash statistic \citep{1979ApJ...228..939C}, for data with a Poisson signal and background. In {\scriptsize XSPEC} this statistic is referred to as cstat. Note that pgstat and cstat denote the log likelihood, $\log \left[ \Pr (y|\theta) \right]$, for the respective data sets. For the joint analysis we consider the fit statistic as the sum of the pgstat and cstat statistics. We perform the analysis using {\scriptsize PyXspec}, a python implementation of HEASARC's {\scriptsize XSPEC} 12.8.1g \citep{1996ASPC..101...17A}.

The GBM detectors are calibrated with a relative uncertainty on the order of 10 \% in effective area \citep{2009ExA....24...47B}. In A19 we found that the best-fit parameter values did not change significantly whether or not we allowed for an effective area correction between the GBM detectors. There is no information on the possible difference in effective area calibration between GBM and XRT. However, we expect that the uncertainty is greater between GBM and XRT than between the GBM detectors. We thus keep the relative normalization between the GBM detectors fixed to unity, whereas we introduce a free relative normalization parameter, $\nr = \mathrm{norm}_{\mathrm{XRT}}~ \mathrm{norm}_{\mathrm{GBM}}^{-1}$, between the GBM and XRT data (also referred to as the cross calibration constant). We let this parameter be fit separately in each time bin. 

\subsection{Priors}
We choose most of our priors to be uninformative. For the luminosity, $\Lum$, and the level of dissipation, $\ed$, we choose log-uniform priors. This means that each decade in these parameters correspond to an equal prior probability. For the Lorentz factor, $\Gamma$, we choose a uniform prior.  

Finally, for the effective area correction between the two instruments, $\nr$, we choose a normal distribution prior centered around $1$ with a standard deviation of $0.1$. Thus, we have
\begin{align*}
    \Pr(\log \ed) & = U(0.01,0.4) \\
    \Pr(\log \Lum) & = U(0.1,1000) \\
    \Pr(\Gamma) & = U(50,1000) \\
    \Pr(\nr) & = \mathcal{N}(\mu = 1,\sigma = 0.1).
\end{align*}

\subsection{Absorption}\label{sec:absorption}
When analyzing X-ray data below 2~keV, Galactic as well as intrinsic (extra-Galactic) absorption becomes relevant. We account for this absorption by using the multiplicative {\scriptsize XSPEC} models \textit{tbvarabs} (for Galactic absorption) and \textit{ztbabs} (for intrinsic absorption). We use the values of the solar abundance vector from \cite{2000ApJ...542..914W} and the cross-section values listed in \cite{1996ApJ...465..487V}. We obtain the weighted total Galactic column density, including the molecular component, $N_{\mathrm{H,gal}}$, from the \textit{Swift} $N_{\mathrm{H,tot}}$ online tool\footnote{\url{http://www.swift.ac.uk/analysis/nhtot/index.php}} \citep{2013MNRAS.431..394W}. In the case when the fraction of molecular hydrogen lies in the range of 10-30\% we have replaced the \textit{tbvarabs} model by \textit{tbabs}, where the fraction of molecular hydrogen is fixed to 20\%. This is the case for GRB~080928, GRB~151027A, and GRB~161117A.

We determine the intrinsic absorption by fitting late time PC data with a tbabs*ztbabs*pow model. At late times the XRT data are dominated by afterglow emission. These data are fainter and are thus captured in PC mode. Pure afterglow spectra are typically well described by power laws, and at late times we expect little spectral evolution (\eg ~\citealt{2009ApJ...698...43R}). Thus, this method is a good way to determine the intrinsic absorption in a way that does not introduce any degeneracy between the spectral model we wish to test and the absorption. For each burst we create a time averaged spectrum consisting of as late PC data as possible. In order to avoid spectral evolution we also require the light curve at these times to be well described by a power law without any breaks. We use the light curve fits from the online catalogue\footnote{\url{http://www.swift.ac.uk/xrt_live_cat/}} to choose these time intervals, which we present in Table~\ref{tab:DataSample}. The fitting to determine $N_{\mathrm{H,intr}}$ is performed using {\scriptsize XSPEC}'s native Maximum-Likelihood scheme. 

Both $N_{\mathrm{H,gal}}$ and $N_{\mathrm{H,intr}}$ are assumed constant for the duration of the burst and kept fixed in all fits with the DREAM model. In Table~\ref{tab:DataSample} we summarize the values of $N_{\mathrm{H,gal}}$ and $N_{\mathrm{H,intr}}$ used. Comparing to the $N_{\mathrm{H,intr}}$ distributions of \cite{2012MNRAS.421.1697C}, we note that the value for GRB~100728A is at the high end of the distribution. Additionally, for GRB~100814A, we find values of $N_{\mathrm{H,intr}}$ consistent with 0.

\subsection{Posterior predictive checks}\label{sec:ppp}
Since we are performing a Bayesian analysis we can use posterior predictive checks (PPCs; \citealt{Gelman96posteriorpredictive,meng1994,doi:10.1177/0049124103257303,Gelman:2013wf}) to assess the quality of our fits. We draw replicated data from the posterior predictive distribution (PPD) using the {\scriptsize XSPEC} command `fakeit'. We then use this new data to assess the quality of our spectral fits. The PPD is the probability of observing some replicated data, conditioned on the observed data, and can be written as
\begin{align*}
    p(y^{\mathrm{rep}} | y) = \int p(y^{\mathrm{rep}}|\theta) p(\theta|y) \ud \theta,
\end{align*}
where $\theta$, $y$, and $y^{\mathrm{rep}}$ are the model parameters, observed data, and replicated data, respectively. $p(\theta|y)$ is the posterior from which we sample using MultiNest. Thus it is easy to construct a posterior predictive $p$-value ($ppp$-value),
\begin{align*} 
    p_{\mathrm{b}} = \mathrm{Pr}\left( T(y^{\mathrm{rep}}) > T(y) | y \right),
\end{align*}
which corresponds to the classical $p$-value averaged over the posterior, $p(\theta|y)$, and where $T$ is a test statistic \citep{10.2307/2240995}. 
We let $T$ be the fit statistic and calculate $p_\mathrm{b}$ for each fit based on $1000$ realizations from the PPD. We consider a fit rejected if $\pb<0.05$.

We stress that an accepted fit only indicates that we cannot reject the fit at the given significance level. It does not mean that the model necessarily can fully describe the data. Additionally, the $ppp$-value of a rejected fit does not tell us how or where the model fails to describe the data. There are many variants of PPCs which can be used to assess the model fitness. We complement our current choice of PPC with manual inspection of posteriors and fits.

\section{Results}\label{sec:results}
\begin{figure*}
    \centering
    \includegraphics[scale=0.42]{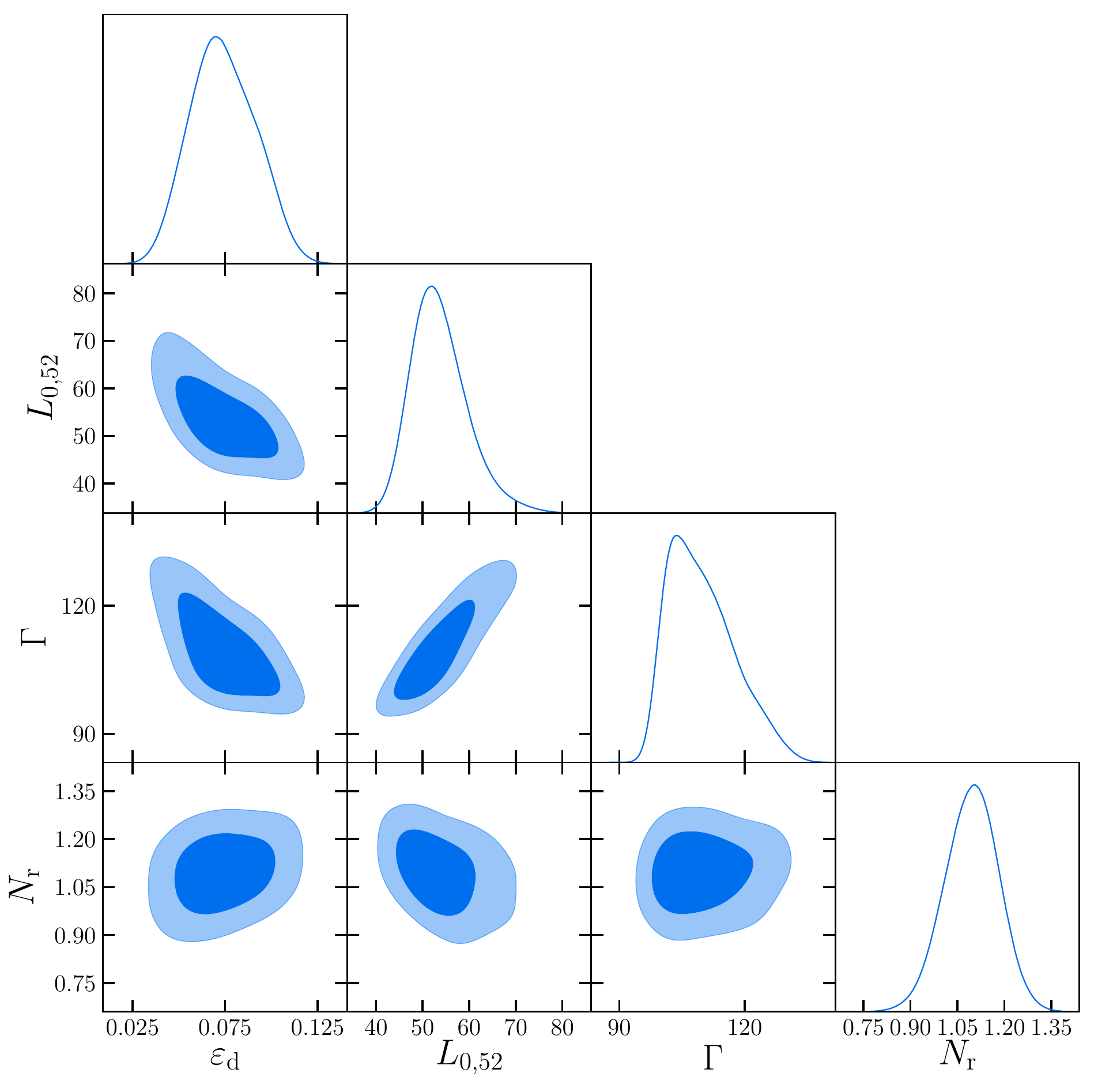}
    \includegraphics[scale=0.42]{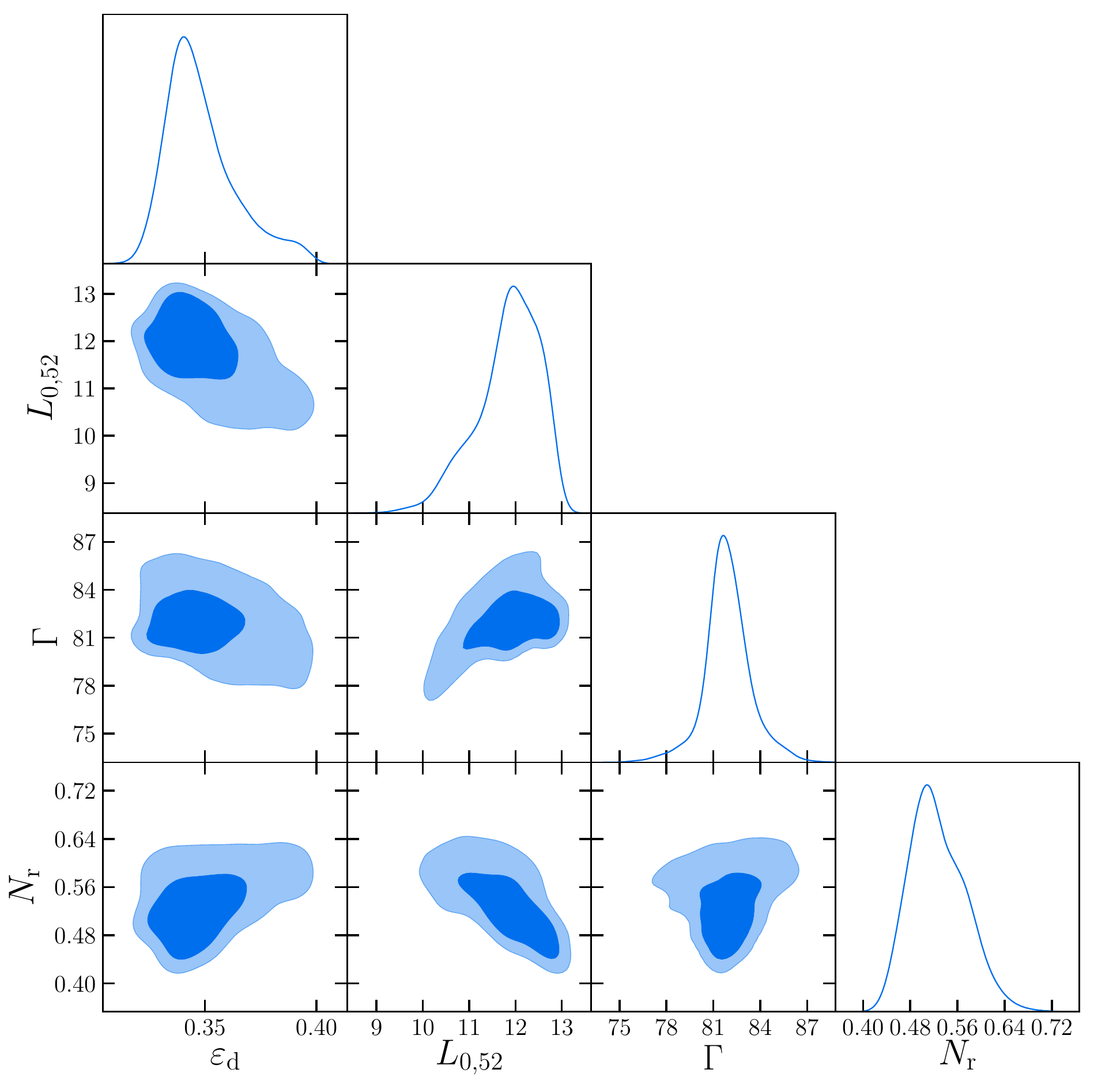}
    \caption{Corner plots showing the full posterior for GRB~100906A at $ 96.8$ - $100.5 $~s and GRB~080928 at $-1.0$ - $12.6$~s, in the left and right panel, respectively. The former passed the PPC, whereas the latter did not. The different contours show the highest posterior density (HPD) regions corresponding to 68 and 95 \% of the probability volume of the joint posterior distributions. 
    The axes of all corner plots are scaled to show the part of the parameter space where the posterior is non-zero. The plots were generated using the python package {\scriptsize getdist}. The complete figure set of corner plots for all analyzed spectra (32 images) is available in the online journal.}
    \label{fig:res:cornerPlot}
\end{figure*}

\begin{figure*}
    \centering
    \includegraphics[scale=0.36]{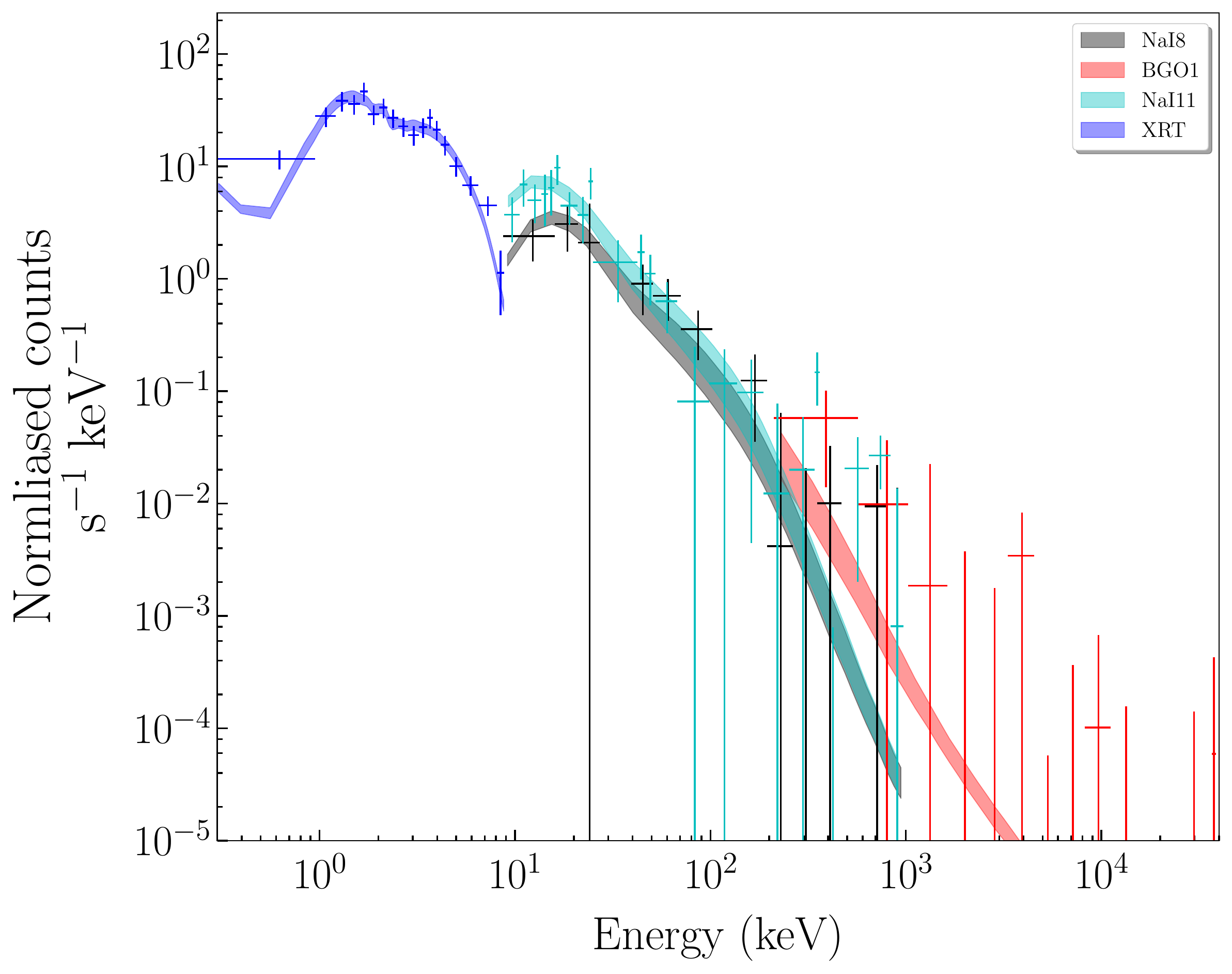}
    \includegraphics[scale=0.36]{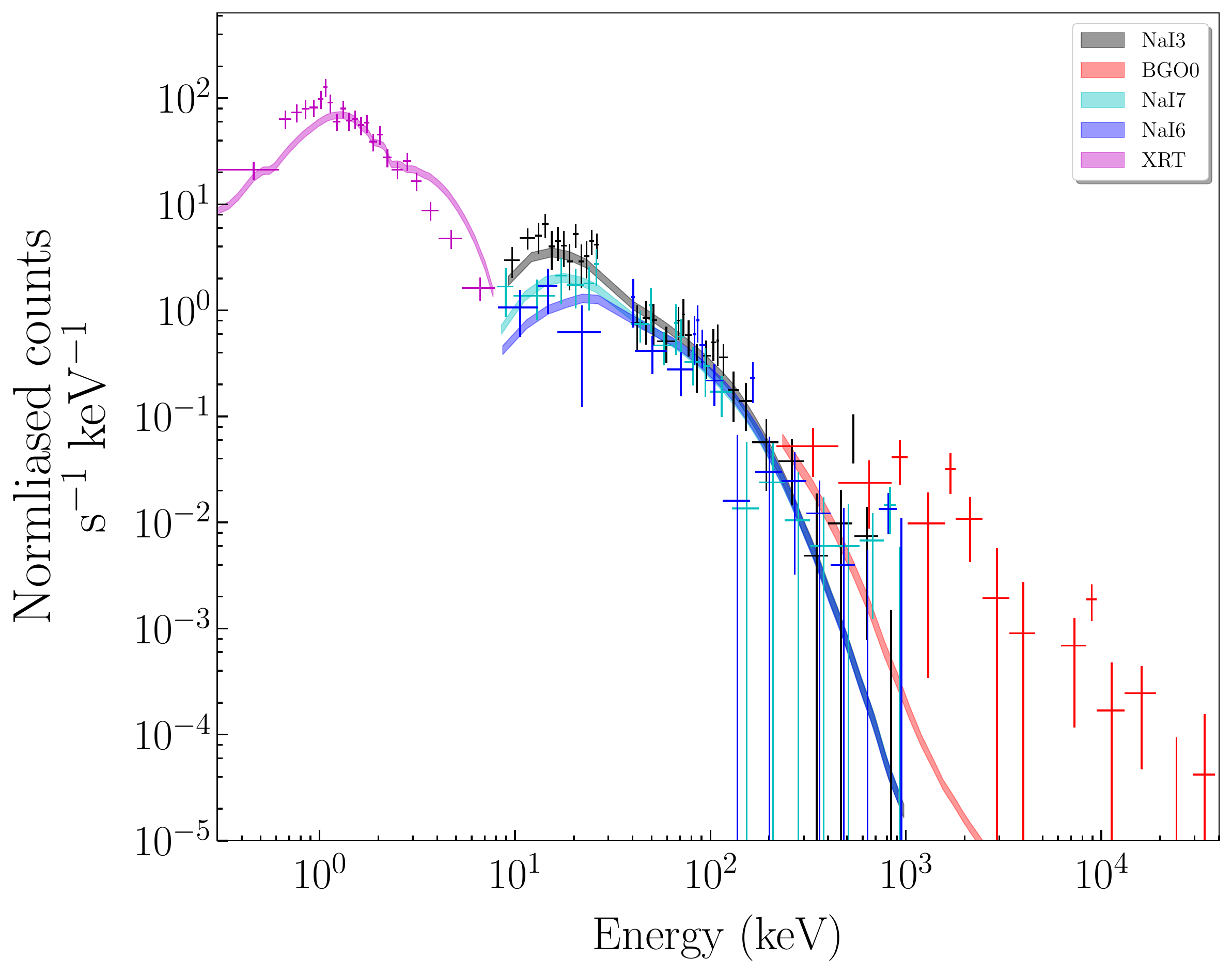}
    \caption{95\% point-wise credible bands of the model shown together with observed counts for GRB~100906A at $ 96.8$ - $100.5 $~s and GRB~080928 at $-1.0$ - $12.6$~s, in the left and right panel, respectively. These are the fits corresponding to the corner plots in Fig.~\ref{fig:res:cornerPlot}. Note that the fit in the left panel passed the PPC whereas the fit in the right panel did not. The legend shows which credible band corresponds to which detector. The data have been visually re-binned to 2$\sigma$ in the GBM data and $5\sigma$ in the XRT. The complete figure set for all accepted fits (16 images) is available in the online journal.
    }
    \label{fig:res:FitPlots}
\end{figure*}

In this section we present the results of the Bayesian analysis described in Section~\ref{sec:analysis}. When performing Bayesian inference we prefer to consider the posterior in its entirety. However, it is often convenient to also use point estimates, here particularly when comparing to the results of A19 and to give an overview of the results in a table. Thus, for point estimates we use the mean of the marginalized posterior for the parameter in question. 
The associated uncertainties correspond to the 1$\sigma$ credible interval centered around the mean, symmetrical in terms of cumulative likelihood. 
In Fig.~\ref{fig:res:cornerPlot} we present examples of corner plots from an accepted and rejected fit, respectively. In Fig.~\ref{fig:res:FitPlots} we show the corresponding fits in data space. Corner plots showing the full posterior of all fits are available as online material. 
Additionally, online we also provide plots corresponding to those in Fig.~\ref{fig:res:FitPlots} for all accepted fits.

\subsection{Accepted fits and parameter estimates}\label{sec:acceptedFits}
A total of 32 time bins were analyzed (17 spectra were not analyzed due to SNR$<3$ in the GBM), with 21 bins being accepted under the posterior predictive checks presented in Section~\ref{sec:ppp}. In Table~\ref{tab:results} we summarize the number of accepted bins for each burst in our sample. We also provide an additional column of spectra which are accepted after further examination of the posterior and fits. This leaves a total of $16$ accepted spectra, as presented further in Section~\ref{sec:XRTimpact}. In Table~\ref{tab:results:parametersEstimates}, we present the point estimates of all fits accepted under the PPC. 
We note that GRB~140512A, GRB~151027A, and GRB~161117A have at least half of the analyzed time bins accepted while having more than 1 analyzed spectrum (see Fig.~\ref{fig:countRateLCs}). 

\begin{table*}[]
    \centering
    \begin{tabular}{ccccc}
        \hline
        GRB & Analyzed & Accepted (PPC) & Accepted (total) & Accepted (total) \\  & & (GBM+XRT) & (GBM+XRT) & (GBM) \\
        \hline
        080928 & 1  & 0 & 0 & 1 \\
        100728A & 3 & 1 & 1 & 1 \\
        100814A & 1 & 1 & 1 & 1 \\
        100906A & 5 & 4 & 2 & 4 \\
        140206A & 3 & 0 & 0 & 0 \\
        140512A & 5 & 4 & 3 & 5 \\
        151027A & 6 & 5 & 5 & 5 \\
        161117A & 8 & 6 & 4 & 6 \\
        \hline
        $\Sigma$ & 32 & 21 & 16 & 23
    \end{tabular}
    \caption{The number of analyzed time-resolved spectra for each burst, together with the number of bins that passed the PPC as described in Section~\ref{sec:ppp}. The fourth column shows the number of accepted spectra after additional consideration has been taken to the posterior distributions and manual inspection of the fits. The fifth column shows the number of bins which are accepted when we consider GBM data only.}
    \label{tab:results}
\end{table*}

\begin{table*}[]
    \centering
    \begin{tabular}{llccccc}
    \hline
   GRB & \multicolumn{1}{p{1.8cm}}{\centering Time bin \\ (s)} & $\varepsilon_{\mathrm{d}}$ &              $L_{0,52}$ &              $\Gamma$ & \multicolumn{1}{p{1.8cm}}{\centering $r_{\mathrm{d}}$ \\
      ($10^{12}$ cm)} &        $N_{\mathrm{r}}$ \\
  \hline
 100728A &                                        134.2-150.5 &     $0.38_{-0.01}^{+0.01}$ &    $15.7_{-0.6}^{+0.6}$ &  $89.5_{-1.3}^{+1.3}$ &                               $11.0_{-0.4}^{+0.4}$ &  $1.32_{-0.07}^{+0.07}$ \\ \hline
 100814A &                                        147.0-150.9 &     $0.27_{-0.05}^{+0.05}$ &    $11.9_{-1.7}^{+1.4}$ &  $96.9_{-3.8}^{+1.5}$ &                                $6.6_{-0.4}^{+0.5}$ &  $0.98_{-0.09}^{+0.09}$ \\ \hline
 100906A &                                         96.8-100.5 &     $0.07_{-0.02}^{+0.02}$ &    $53.6_{-5.8}^{+5.8}$ &       $110_{-8}^{+8}$ &                               $20.6_{-2.9}^{+2.9}$ &  $1.10_{-0.09}^{+0.08}$ \\
 100906A &                                        100.5-105.0 &     $0.09_{-0.02}^{+0.02}$ &    $72.5_{-7.2}^{+7.0}$ &       $114_{-9}^{+8}$ &                               $25.2_{-3.8}^{+4.0}$ &  $1.00_{-0.08}^{+0.08}$ \\
 100906A &                                        105.0-110.0 &     $0.08_{-0.02}^{+0.02}$ &    $74.8_{-7.5}^{+8.0}$ &       $113_{-9}^{+9}$ &                               $26.4_{-4.0}^{+4.2}$ &  $1.12_{-0.08}^{+0.08}$ \\
 100906A &                                        110.0-115.4 &     $0.02_{-0.01}^{+0.01}$ &  $89.3_{-10.0}^{+10.3}$ &       $113_{-9}^{+9}$ &                               $31.5_{-4.6}^{+4.7}$ &  $1.10_{-0.08}^{+0.08}$ \\ \hline
 140512A &                                        108.9-114.2 &     $0.37_{-0.02}^{+0.02}$ &     $1.7_{-0.1}^{+0.1}$ &  $95.5_{-1.2}^{+1.2}$ &                                $1.0_{-0.0}^{+0.0}$ &  $0.99_{-0.09}^{+0.09}$ \\
 140512A &                                        128.2-142.6 &     $0.40_{-0.00}^{+0.00}$ &     $3.6_{-0.1}^{+0.1}$ &  $95.8_{-0.4}^{+0.4}$ &                                $2.1_{-0.0}^{+0.0}$ &  $1.19_{-0.05}^{+0.05}$ \\
 140512A &                                        142.6-145.6 &     $0.28_{-0.05}^{+0.05}$ &     $1.4_{-0.2}^{+0.2}$ &  $82.9_{-5.0}^{+5.1}$ &                                $1.3_{-0.1}^{+0.1}$ &  $0.96_{-0.09}^{+0.09}$ \\
 140512A &                                        145.6-156.0 &     $0.22_{-0.02}^{+0.02}$ &     $1.2_{-0.1}^{+0.1}$ &  $86.2_{-2.0}^{+2.2}$ &                                $0.9_{-0.0}^{+0.0}$ &  $1.04_{-0.08}^{+0.08}$ \\ \hline
 151027A &                                         97.1-105.7 &     $0.31_{-0.04}^{+0.04}$ &     $1.6_{-0.2}^{+0.2}$ &  $81.6_{-3.7}^{+3.6}$ &                                $1.5_{-0.1}^{+0.1}$ &  $1.09_{-0.08}^{+0.08}$ \\
 151027A &                                        105.7-109.1 &     $0.39_{-0.01}^{+0.01}$ &     $4.2_{-0.2}^{+0.2}$ &  $94.6_{-1.0}^{+1.0}$ &                                $2.5_{-0.1}^{+0.1}$ &  $1.05_{-0.07}^{+0.07}$ \\
 151027A &                                        109.1-112.1 &     $0.38_{-0.01}^{+0.01}$ &     $5.6_{-0.2}^{+0.2}$ &  $94.7_{-1.0}^{+1.0}$ &                                $3.3_{-0.1}^{+0.1}$ &  $1.07_{-0.07}^{+0.08}$ \\
 151027A &                                        112.1-115.2 &     $0.35_{-0.04}^{+0.04}$ &     $4.5_{-0.4}^{+0.4}$ &  $86.5_{-2.9}^{+2.8}$ &                                $3.5_{-0.2}^{+0.2}$ &  $1.14_{-0.07}^{+0.08}$ \\
 151027A &                                        115.2-120.0 &     $0.21_{-0.04}^{+0.04}$ &     $3.2_{-0.5}^{+0.4}$ &  $77.5_{-6.1}^{+5.0}$ &                                $3.5_{-0.3}^{+0.3}$ &  $1.20_{-0.08}^{+0.08}$ \\ \hline
 161117A &                                          72.5-90.2 &     $0.07_{-0.01}^{+0.01}$ &    $20.1_{-2.1}^{+2.3}$ &      $153_{-7}^{+10}$ &                                $2.8_{-0.2}^{+0.2}$ &  $0.93_{-0.08}^{+0.08}$ \\
 161117A &                                          90.2-92.7 &     $0.09_{-0.02}^{+0.02}$ &    $35.7_{-5.4}^{+5.6}$ &     $162_{-13}^{+13}$ &                                $4.2_{-0.5}^{+0.4}$ &  $1.00_{-0.09}^{+0.09}$ \\
 161117A &                                         92.7-100.2 &     $0.07_{-0.01}^{+0.01}$ &    $35.5_{-2.5}^{+2.8}$ &       $137_{-4}^{+4}$ &                                $7.0_{-0.4}^{+0.4}$ &  $1.08_{-0.08}^{+0.09}$ \\
 161117A &                                        100.2-104.1 &     $0.08_{-0.01}^{+0.01}$ &    $52.2_{-4.0}^{+4.6}$ &       $134_{-5}^{+6}$ &                               $10.9_{-0.9}^{+0.7}$ &  $1.05_{-0.09}^{+0.09}$ \\
 161117A &                                        131.0-138.5 &     $0.09_{-0.01}^{+0.01}$ &    $36.0_{-2.2}^{+2.2}$ &  $97.4_{-0.8}^{+0.8}$ &                               $19.6_{-1.1}^{+1.1}$ &  $1.01_{-0.07}^{+0.07}$ \\
 161117A &                                        138.5-144.2 &     $0.05_{-0.01}^{+0.01}$ &    $26.9_{-2.4}^{+2.5}$ &  $94.1_{-1.2}^{+1.3}$ &                               $16.2_{-1.3}^{+1.3}$ &  $0.94_{-0.09}^{+0.09}$ \\ 
  \hline
 
    \end{tabular}
    \caption{Point estimates of model parameters and $\rd$ for all 21 fits which passed the PPC in our sample. We also include values of $\rd$, which is not one of the fit parameters, but obtained from the relation $\rd = r_{\mathrm{ph}}/\tau = L \sigma_{\mathrm{T}} / 4\pi \tau \Gamma^3  c^3 m_{\mathrm{p}}$, as given in Section~\ref{sec:physscenario}}
    \label{tab:results:parametersEstimates}
\end{table*}

\begin{figure}
    \centering
    \includegraphics[scale=0.53]{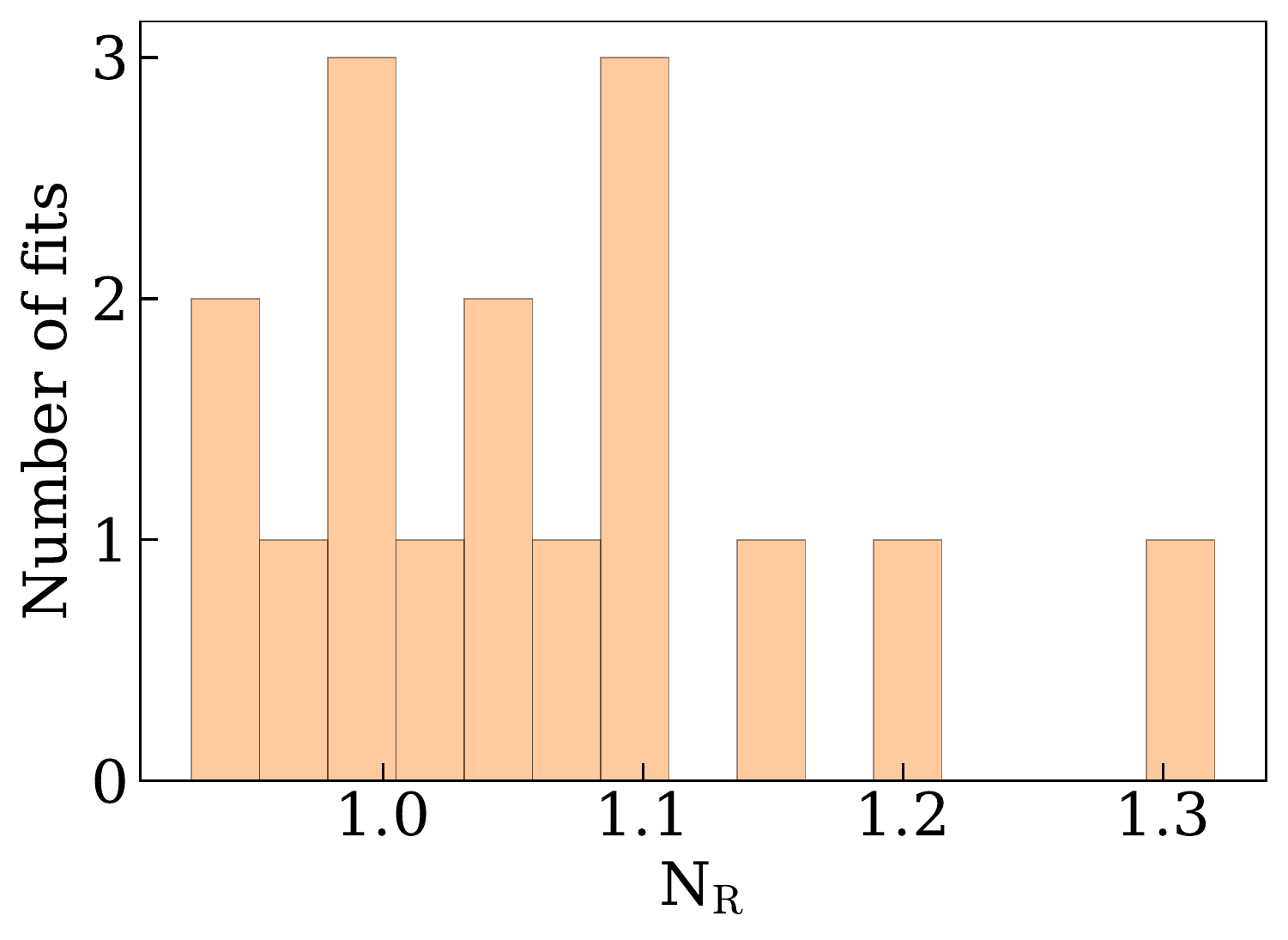}
    \caption{Histogram of the cross calibration constant. The histogram includes the 21 fits which have passed the PPC.}
    \label{fig:XRTRelNormDistribution}
\end{figure}

In Fig.~\ref{fig:XRTRelNormDistribution} we show the distribution of the relative normalization parameter, $\nr$. We note that these values are generally larger than unity, as discussed further in Section~\ref{sec:XRTimpact} below. There are too few analyzed and accepted bins to make any reliable inferences about the temporal evolution. However, as can be seen in Table~\ref{tab:results:parametersEstimates}, the parameter estimates do not vary erratically throughout a burst. Additionally, we note that the parameters evolve according to the trends we observed in A19. There the most prominent trends were found to be that $\Lum$ follows the light curve.

\subsection{The impact of XRT data}\label{sec:XRTimpact}
\begin{figure}
    \centering
    \includegraphics[scale=0.55]{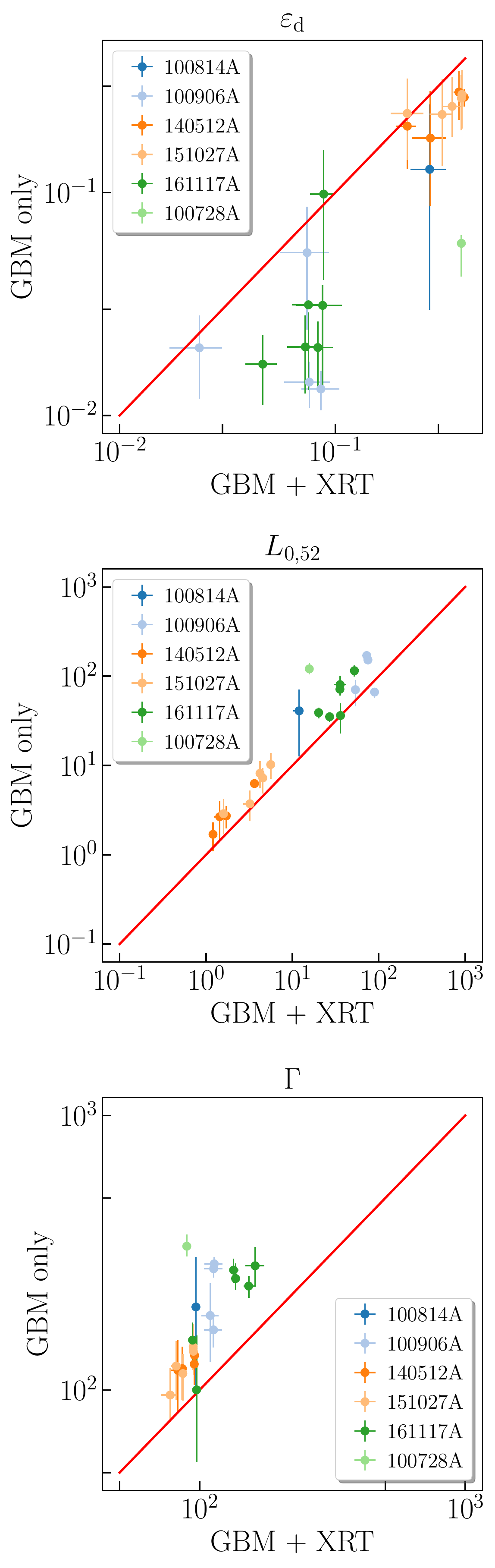}
    \caption{Comparison of parameter estimates from the analysis with and without XRT data. We show parameter estimates of the three free model parameters, $\ed$, $\Lum$ and $\Gamma$ in the top, middle, and bottom panel, respectively. The red line corresponds to a $1:1$ relation. The figure only includes data points from fits that passed the PPC. The error bars represent the 1$\sigma$ uncertainties.}
    \label{fig:compXRTvsGBMXRT}
\end{figure}

To examine how the XRT data affect the fits we also perform the analysis with the GBM data only. In Fig.~\ref{fig:compXRTvsGBMXRT} we show how parameter estimates change depending on whether we include XRT data in the analysis. It is clear that there are systematic differences in all parameters, although it is most striking for $\ed$ and $\Gamma$. The effect is especially prominent for $\Gamma$, with all estimates of $\Gamma$ decreasing as we introduce XRT data. For $\ed$ we note that there is a tendency that estimates are higher when we include XRT data. We stress that the parameter estimates with and without XRT-data are still consistent within 3$\sigma$ uncertainties in the majority of cases.

\begin{figure*}
    \centering
    \includegraphics[scale=0.6]{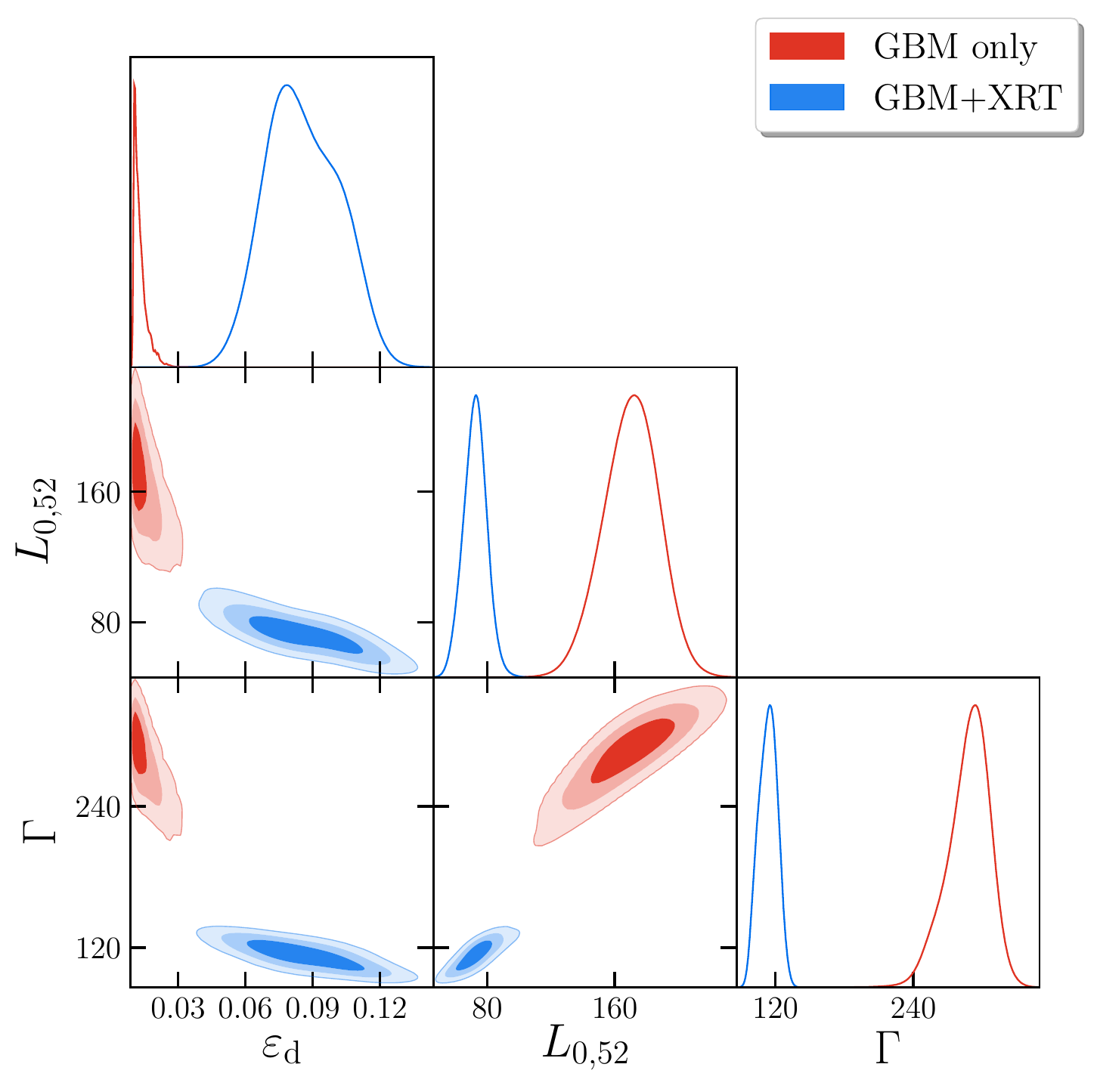}
    \includegraphics[scale=0.6]{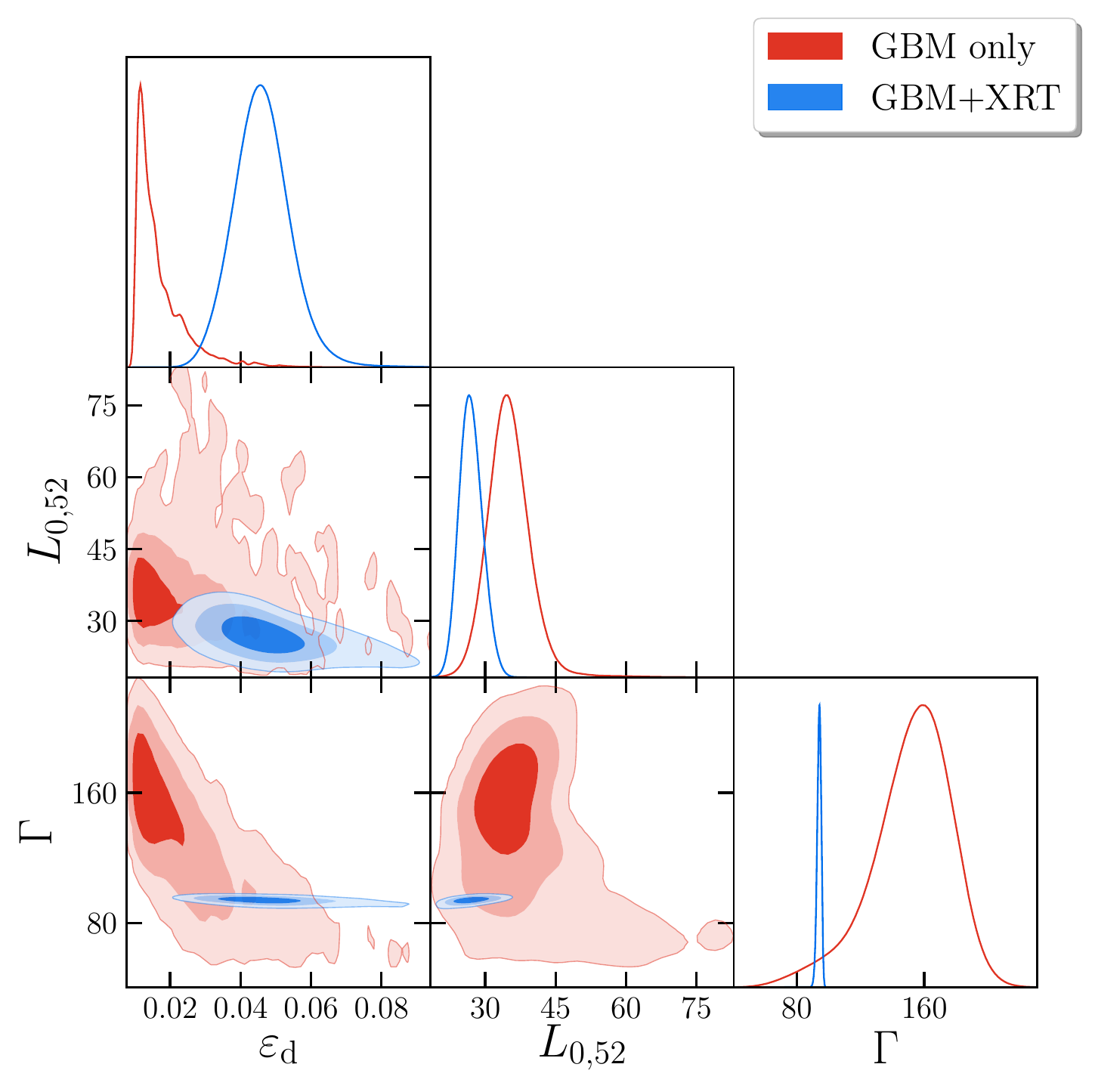}
    \caption{Posteriors for the joint GBM-XRT analysis and the analysis of GBM data only in blue and red, respectively. The left and right panels show the case of GRB~100906A at $100.5$ - $105.0$~s (the third time bin) and GRB~161117A at $138.5$-$144.2$~s (the eighth time bin), respectively. The contours denote the 68, 95 and 99.7\% HPD regions. The fit in the left panel is rejected manually due to the disjunct posterior distributions, whereas the fit in the right panel shows an accepted fit. The complete figure set for all fits that passed the PPC (21 images) is available in the online journal.
    }
    \label{fig:compPosteriors}
\end{figure*}

However, $5$ fits remain with non-overlapping posteriors at this level. The spectra in question are GRB~100906A (time bins $3$ and $4$), GRB~140512A (time bin $3$) and GRB~161117A (time bins $3$ and $4$). This indicates that the model cannot describe these data adequately. These fits are not rejected by the PPC since it only considers the overall spectral shape, and not the consistency of different subsets of the data. In Fig.~\ref{fig:compPosteriors} we show an example of these non-overlapping posteriors for a spectrum from GRB~100906A. The fact that the posteriors in the left panel are disjunct indicates that the joint fit to GBM-XRT is rejected by the fit in the GBM energy range, and vice versa.
We therefore treat these spectra as rejected, even though they passed the PPC. These fits are marked in yellow in Fig.~\ref{fig:countRateLCs}. We comment on this further in section~\ref{sec:discuss:XRTimplications:AnalysisUncertainties}. In Fig.~\ref{fig:compPosteriors} we also show an example of posteriors that is more representative of the accepted sample as a whole, where there is significant overlap of the posteriors. Additionally, this figure illustrates that the quality of the constraints improve significantly when XRT data are added. This is a result of both the increased energy range and the quality of the XRT data.

In Table~\ref{tab:results}, we also present the number of fits which passed the PPC when we do not include the XRT data. 
Not surprisingly, we note that we obtain additional accepted fits when we remove the XRT data (2 extra spectra pass the PPC). This is because the model becomes less constrained, making it harder to reject the fits. 
However, most fits remain accepted when XRT data are introduced, indicating that the model is overall consistent with the XRT data.
It is no surprise that there are spectra which are poorly described by the model. As already found based on fits to GBM data in A19, GRB~100728A and GRB~140206 are not well described by the model.

\begin{figure*}
    \centering
    \includegraphics[scale=0.35]{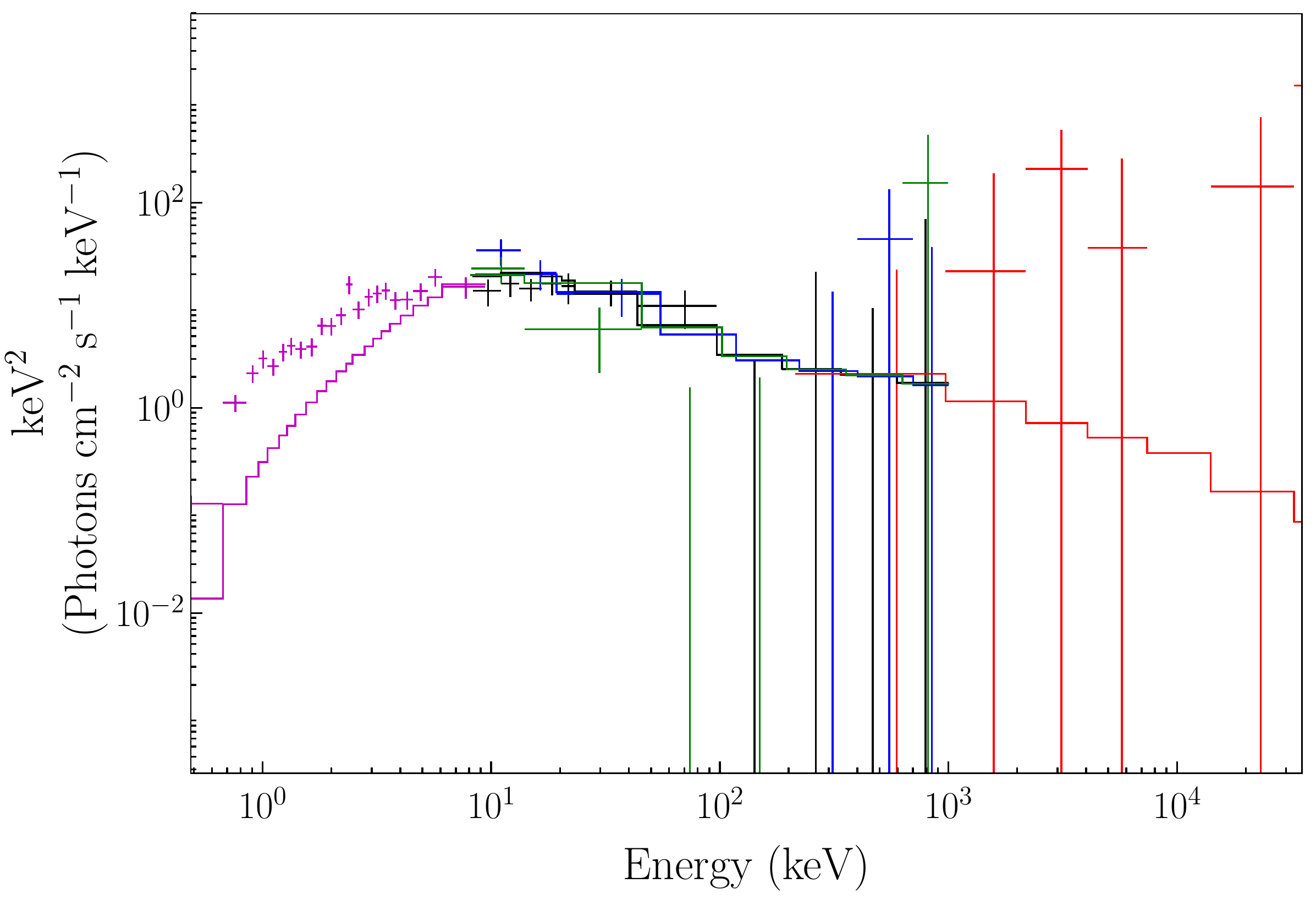}
    \includegraphics[scale=0.35]{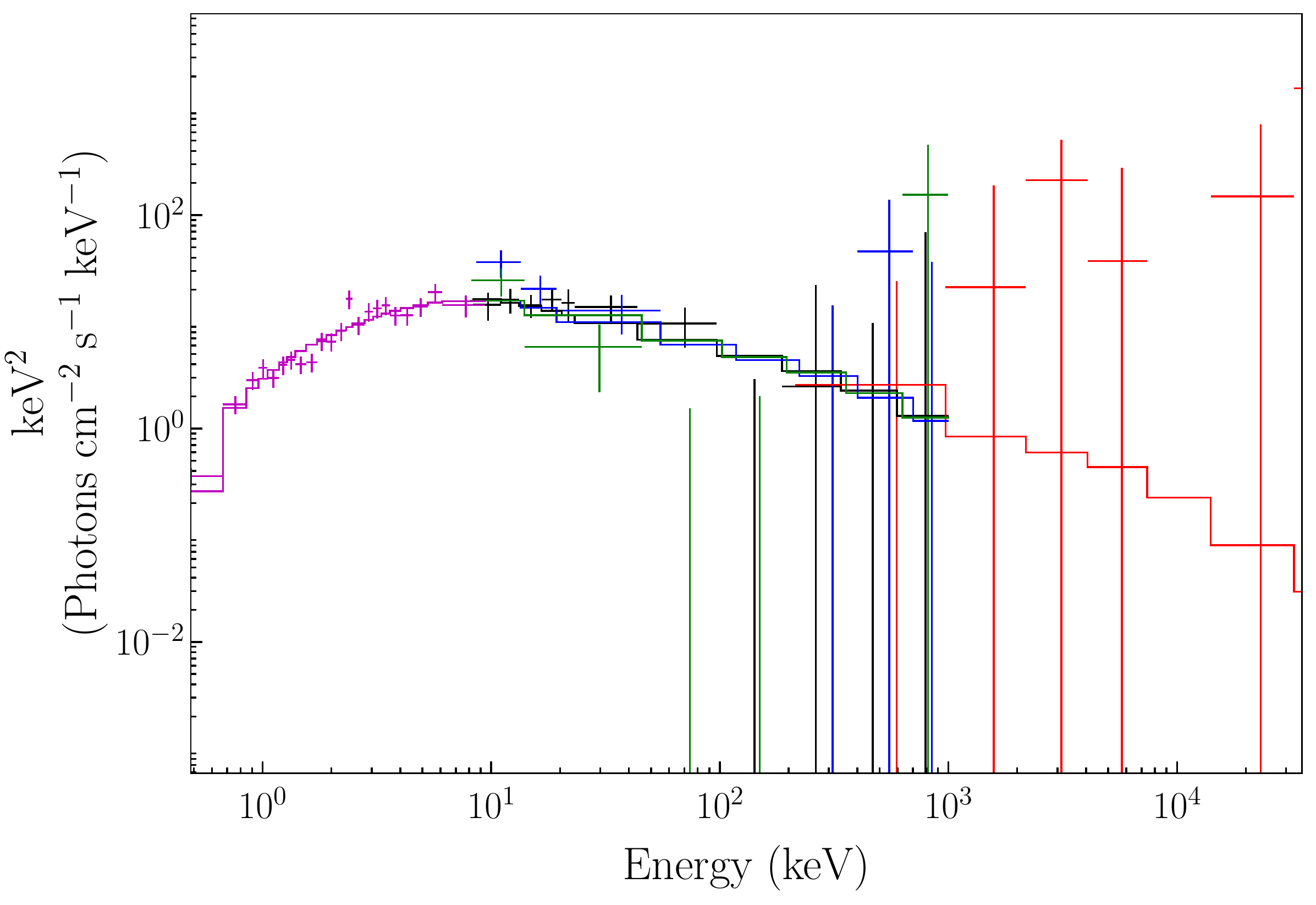}
    \caption{Examples of best fit models to GRB~161117A at $138.5$-$144.2$~s. The magenta, black, blue, green and red colours represent data and model for the XRT, NaI2, NaI1, NaI10 and BGO0 detectors, respectively. The left panel shows the result of fitting DREAM1.3 to only the GBM data, with the XRT data being added to the figure without having been used in the analysis. Note that the extrapolation of the fit down to $0.3$~keV includes absorption. The right panel shows the fit to all data. The parameters are $\ed = 0.017$, $\Lum = 35.2$, $\Gamma = 152$, and $\ed = 0.046$, $\Lum = 26.9 $, $\Gamma = 94.1$ in the left and right plot, respectively.
    Our PPC accepts both the fit in the left (without the XRT data) and right figure, respectively. Note that plotting only the best-fit model in $\nu F_\nu$ is not the best representation of either the data or the posterior. This representation is chosen here to provide a qualitative comparison with model spectra for the discussion in Section~\ref{sec:discuss:XRT}.}
    \label{fig:res:compareFits_regview}
\end{figure*}

In order to illustrate why the parameters change when XRT data are added, we show in Fig.~\ref{fig:res:compareFits_regview} the best fit model to the eighth time bin of GRB~161117A when the analysis has been performed with and without the XRT data (right and left panels, respectively). We let the best fit be represented by the model at the posterior mean. 
In this example the best-fit model for the GBM data clearly under-predicts the XRT data (left panel of Fig.~\ref{fig:res:compareFits_regview}). We note that the posteriors have significant overlap within the 3$\sigma$ credible region (right panel of Fig.~\ref{fig:compPosteriors}).

Further, we find a similar excess of counts at low energy in the majority of spectra. This is consistent with our finding of the cross calibration constant generally being larger than unity (see Fig.~\ref{fig:XRTRelNormDistribution}). This is because the cross calibration constant is often able to account for some of this excess. For the remaining counts, the logic of the change in parameters is that $\Gamma$ decreases in order to accommodate the need for additional low-energy counts, which shifts the spectrum to lower energies. In order to maintain the spectral peak, $\ed$ increases. $\Lum$ decreases to adjust the low energy slope and to preserve the total energy in the spectrum.

\section{Discussion}\label{sec:discussion}
We have studied a specific version of the subphotospheric dissipation model assuming localized dissipation due to internal shocks ($r_{\rm d} = \Gamma^2 r_0$) in a flow with negligible magnetization. 
In A19 we found that this model was unable to describe the brightest part of the GRB population. Since characteristic features in the model occur below the GBM energy range, we have in this paper included data in the soft X-ray range in order to further constrain the model. We find that 50\% of the analyzed spectra can be described by the model. Below we discuss these results.

We begin the discussion by considering possible uncertainties in the analysis and the impact these may have on the results (section~\ref{sec:discuss:XRTimplications:AnalysisUncertainties}). This also includes a discussion of the limitations of the PPC and additional evaluation of the fits. We then examine what differentiates accepted and rejected fits (section~\ref{sec:discussion:accepted}). This is followed by a comparison to the results of A19 (section~\ref{sec:discussion:paper2}). We also discuss the possibility of additional emission components in the XRT data (section~\ref{sec:discuss:XRTimplications:AdditionalEmissionComponents}).
Finally, we discuss the implications of these results for DREAM, and what changes to the subphotospheric dissipation scenario can be made to accommodate this (section~\ref{sec:discuss:XRTimplications:modelImplications}).

\subsection{Uncertainties in the analysis}
\label{sec:discuss:XRTimplications:AnalysisUncertainties}
The main uncertainties in the analysis are the relative normalization, $\nr$, and the intrinsic absorption, $N_{\mathrm{H,intr}}$. We have investigated the effect of allowing for larger values of $\nr$ by changing its prior to $\mathcal{N}(\mu=1,\sigma = 0.5)$. Only $6$ of $32$ spectra have a marginalised posterior mean for $\nr$ which changes more than 10\% when we use this prior. Furthermore, \cite{2017ApJ...846..137O} find that the relative calibration between \textit{Swift} and GBM data agrees within 15\%. This is compatible with our original prior, and supports the conclusion that the relative calibration differences have been adequately accounted for. Additionally, tests show that the results are not sensitive to our choice of priors for the other parameters.

We also examined the sensitivity of the results to the value of $N_\mathrm{H,intr}$. We did this by performing the analysis using the 1$\sigma$ lower and upper bounds on $N_{\mathrm{H,intr}}$ from Table~\ref{tab:DataSample}. Using the lower bounds of $N_{\mathrm{H,intr}}$, only $1$ fit change any of its parameter estimates (\ie ~the mean of the corresponding marginalized posterior) by more than $10\%$. Similarly, only $2$ fits change when using the upper bounds. These spectra are GRB~080928 at $-0.982$ - $12.637$~s for high absorption and GRB~100906A at $115.367$ - $121.375$~s for both high and low absorption. Both of these spectra are rejected in the original analysis. This indicates that our results are also robust to uncertainties in $N_{\mathrm{H,intr}}$.

Using Bayesian inference, information about degeneracies in the model fits is available through the posteriors. Inspecting the corner plots of all fits, we note that there is a tendency to a $\Gamma$ - $\ed$ degeneracy. However, it is sufficiently weak to indicate that the changes in $\Gamma$ and $\ed$ that we observe in Fig.~\ref{fig:compXRTvsGBMXRT} when including XRT data are not caused primarily by model degeneracies. However, there is a slightly stronger degeneracy present in $\Gamma$ - $\Lum$. This is interesting given the $\Gamma$ - $\Lum$ correlation found in A19. Upon closer inspection we find that this degeneracy is generally small and it was demonstrated in A19 that the correlation is not caused by model degeneracies. There is also a weak degeneracy between $\ed$ - $\Lum$. This is not surprising given that $\ed \Lum$ sets the total energy in the spectrum, which is expected to remain constant. Notably, we find no degeneracies between the cross calibration constant, $\nr$, and the other fit parameters.

Regarding the PPC, the chosen $ppp$-value threshold of $\pb > 0.05$ will naturally affect the number of accepted spectra. However, we note that the number of spectra do not vary much for small alterations of $\pb$, with 21 and 22 accepted spectra for $\pb > 0.1$ and $\pb > 0.01$, respectively (compared to the 21 accepted spectra at $\pb > 0.05$). Thus, despite the small number of analyzed and accepted spectra, our conclusions are not particularly sensitive to the choice of threshold value in the PPC. 

As noted in Section~\ref{sec:XRTimpact}, $5$ spectra exhibit significant ($> 3\sigma$) inconsistencies between the posteriors from the joint GBM-XRT fits and the GBM data only fits (indicated with yellow shading in Fig.~\ref{fig:countRateLCs}).
Manual inspection shows that these fits deteriorate when XRT data are added, but not enough to cause the PPC to reject the fits. 
However, the posteriors are sufficiently inconsistent for us to label these fits as rejected for the purpose of this analysis. 
The fact that the fits still pass the PPC is likely a result of the model's flexibility and the relatively low SNR of the GBM data. This may also be affected by calibration uncertainties.

\subsection{Accepted and rejected fits}\label{sec:discussion:accepted}
\begin{figure}
    \centering
    \includegraphics[scale=0.35]{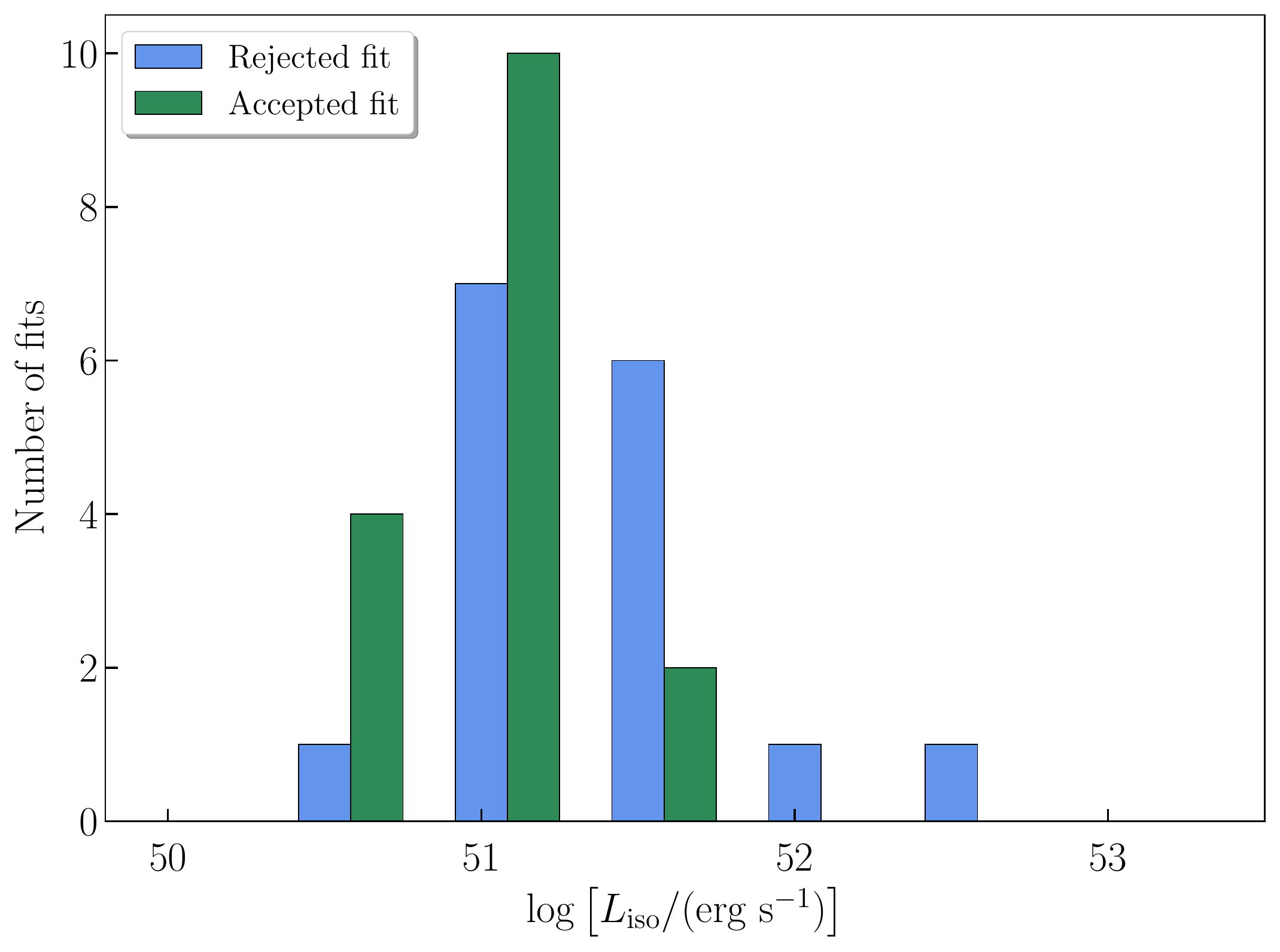}
    \caption{Histograms of the logarithm of the isotropic equivalent luminosity, $L_{\mathrm{iso}}$. We show two bins per decade, spaced uniformly. The bars have been separated for visual clarity. Blue and green bars indicate fits which failed and passed the PPC, respectively. $L_\mathrm{iso}$ was obtained from fits with a cut-off power law.}
    \label{fig:discussion:LisoHistogram}
\end{figure}

\begin{figure}
    \centering
    \includegraphics[scale=0.35]{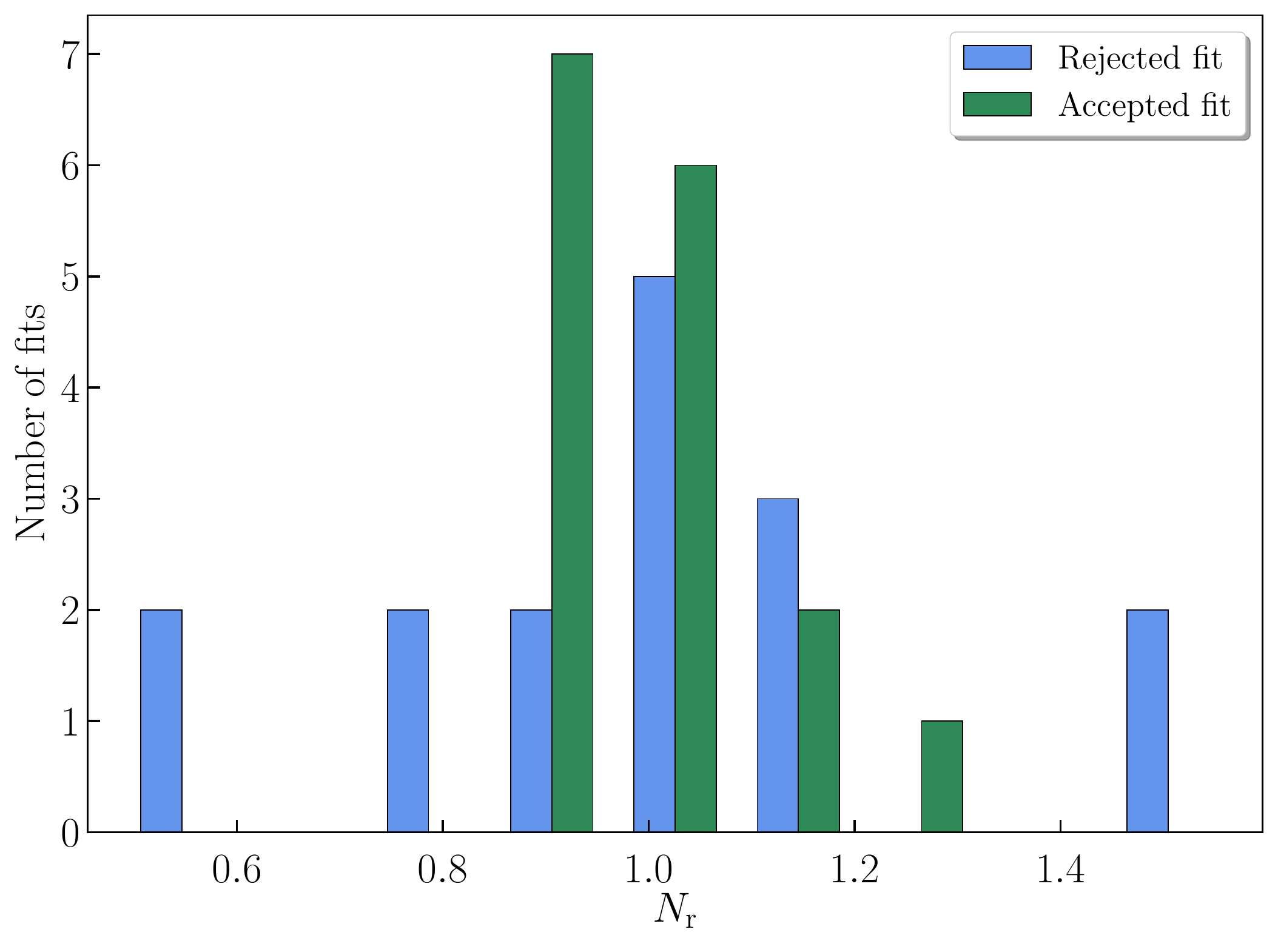}
    \caption{Histograms of the cross calibration constant $\nr$. Blue and green bars indicate fits which failed and passed the PPC, respectively. The bars have been separated for visual clarity.}
    \label{fig:discussion:NrHistogram}
\end{figure}

In order to investigate why the model fails to describe about half of the analyzed spectra we search for systematic trends between whether a fit is rejected and the properties of the data. We find only weak correlations between if a fit is rejected and its corresponding parameter estimates. As shown in Fig.~\ref{fig:discussion:LisoHistogram}, there is a tendency of rejecting fits as the model luminosity becomes larger. It also appears that fits with the most extreme values of the cross calibration constant, $\nr$, are rejected (see Fig.~\ref{fig:discussion:NrHistogram}). Although there are not enough data points to ascertain any statistically significant correlations, the luminosity relation is well-known from A19. Furthermore, a cross calibration constant that deviates significantly from unity suggests that the model cannot adequately describe the data and is compensating for it by improbable values of the artificial normalization parameter. Thus, these correlations are not surprising.
There is also an indication that the rejected fits coincide with the peaks of the light curves (see Fig.~\ref{fig:countRateLCs}). This is likely a result of the model's difficulty in describing spectra with a high luminosity. It could also indicate that the XRT peak flux is dominated by another emission mechanism (discussed further in Section~ \ref{sec:discuss:XRTimplications:AdditionalEmissionComponents}). However, as that this trend is not clear for all bursts, it may simply be a result of the small number of bins in the analysis.

We find no significant correlation between if a fit passes the PPC and the observed flux in the GBM or XRT, the SNR in any detector, or any of the model parameters obtained when fitting with a power law and cutoff-power law for the XRT and GBM data, respectively. Due to the small number of analyzed bins we cannot draw any strong conclusions from this.

\subsection{Comparison to A19}\label{sec:discussion:paper2}
The DREAM model was initially introduced in A15 where it was fit to two GRBs. In A19 we implemented significant improvements to the model and fit it to a larger sample of 36 GRBs.
Here we have further tested the model by including XRT data at low energies in the analysis. This is because of the distinctive model predictions in this range, which result from the assumptions of internal shocks and negligible magnetization. The other main differences from A19 are that we have expanded the parameter space of the model slightly and analyzed the data using Bayesian inference. The model parameter space was extended with several new grid points in $\Gamma$, extending the grid down to $\Gamma = 50$ and up to $\Gamma = 1000$, and one point in $\Lum$, extending it up to $\Lum =1000$. The joint GBM-XRT fits mainly inhabit the low $\Gamma$ part of this extended parameter space.

The higher $\Lum$ and $\Gamma$ would lead to additional accepted fits in the A19 sample. The difference in SNR cut would also lead to a larger number of analyzed spectra. Performing fits with the new model to the GBM data of the GRBs in the current sample primarily affects GRB~100728A and GRB~100906A. The fraction of accepted fits increase from $4$ to $44$ \% and from $26$ to $41$ \% in the bright interval before the XRT observations start for GRB~100728A and GRB~100906A, respectively. We note that these GRBs are still not fully described by the model and that the highest $\Lum$ of both bursts push the limit of what is physically plausible.

One of the main conclusions of A19 was that DREAM is unable to account for the brightest GRBs. Fig.~\ref{fig:discussion:LisoHistogram} is reminiscent of the corresponding Fig.~11 in A19, although the peak of the distribution is shifted to lower luminosities in this work (due to the sample in A19 being biased towards high luminosity bursts). However, we do not observe the problem of under-predicting the observed luminosity in this analysis. 
Fig.~\ref{fig:discussion:LisoHistogram} can instead be attributed to the fact that higher luminosities result in less flexibility for the model. It is then harder to describe data which results in high estimates of the luminosity, even when achieving a sufficiently high luminosity is not an issue. This was also discussed in A19. However, as pointed out in Section~\ref{sec:discussion:accepted}, Fig.~\ref{fig:discussion:LisoHistogram} could also be the result of the small number of analyzed spectra.

\subsection{Implications of XRT data}\label{sec:discuss:XRT}
As noted in Section~\ref{sec:XRTimpact}, the inferred parameter estimates change systematically as we add XRT data. However, for a majority of spectra, the changes are within reasonable uncertainties of the fits to the GBM only data. Conversely, five spectra exhibit significant inconsistencies between the posterior of the joint GBM-XRT analysis and the GBM only analysis, and are thus rejected, as discussed in Section~\ref{sec:discuss:XRTimplications:AnalysisUncertainties}. 
The fact that the XRT data make a significant difference in the analysis of prompt emission has also been found by \cite{2017ApJ...846..137O,2018A&A...616A.138O}. It is clear that there are spectra in this sample which the DREAM model cannot describe. This may of course be due to the fact that it simply does not represent the correct dissipation scenario or emission process for these bursts. However, there are also other possibilities that do not necessarily rule out the physical scenario considered here. We discuss these in turn below.

\subsubsection{Presence of additional components}
\label{sec:discuss:XRTimplications:AdditionalEmissionComponents}
It is possible that the emission observed by XRT is of a (partly) different origin than that observed by GBM. Since DREAM is a one-zone model and we consider no additional components in the analysis, this could help explain some of the cases where we are unable to find good fits. 
Additional components can originate from different emission sites within the jet itself (\eg ~an optically thin component), a region outside the jet (\eg ~cocoon emission), or by interactions of the jet with the surrounding medium (\eg ~afterglow). However, we note that all XRT light curves have flares, suggesting that this emission is closely related to prompt emission. Additionally, most light curves exhibit significant similarities in the GBM and XRT light curves (see Fig.~\ref{fig:countRateLCs}).
The main difference that can be seen is that the XRT light curves are often stronger at the end of the bursts, resembling a kind of delay. This may reflect the commonly observed softening of the prompt emission, with the spectral peak moving to lower energies with time. A particularly clear example is GRB 151027A, which has very bright XRT emission peaking $\sim 15$~s later than the GBM light curve. Additionally, GRB~100906A shows a similar delay while also having weak GBM emission in the overlap, leading to a clear difference in the light curve morphology between the XRT and GBM.

GRB~151027A is the only GRB in our sample with an extra component reported in the literature.
\cite{2017A&A...598A..23N} find a blackbody component at soft X-rays from joint fits to GBM, BAT, and XRT data. This was confirmed by \cite{2018MNRAS.474.2401V}, performing an analysis of XRT data using a power law plus a blackbody. We note that the DREAM model provides adequate fits in the early times of XRT data, when the flux from the additional blackbody is low. When the blackbody becomes significant, we instead obtain a poor fit (see \citealt{2017A&A...598A..23N,2018MNRAS.474.2401V} for discussions on the blackbody flux).
\cite{2017A&A...598A..23N} suggests that the blackbody originates from the re-acceleration of the fireball.
Another possibility is that the thermal component originates from a hot cocoon surrounding the jet, as discussed by \cite{2018MNRAS.474.2401V}.
Neither scenario can be described using the DREAM model, since it is a one-zone photospheric model with no hydrodynamical evolution.
Alternatively, an extra blackbody component may indicate dissipation occurring just below the photosphere, which is known to produce double-peaked spectra (see A15). This scenario would not be captured by the current version of the model due to the assumption of dissipation at $\tau = 35$.

Performing the same analysis of the XRT data as in \cite{2018MNRAS.474.2401V}, we find no significant blackbody in any of the other bursts. However, additional components may have different spectral shapes, which could account for the excess seen in some of the GRBs. 
This is unlikely to be the only explanation for all the poor fits though, as the model clearly fails to describe many intervals in GRBs where the light curves are well correlated (\cf Fig.~\ref{fig:countRateLCs}).

\subsubsection{Model implications}
\label{sec:discuss:XRTimplications:modelImplications}
In A19 we argued that a different dissipation scenario than internal shocks is needed for the brightest GRBs. The study presented in this paper has revealed additional issues at soft X-rays in some bursts. 
Assuming no additional components are present, this suggests that the current implementation of the model does not capture all the relevant conditions and processes. Here we discuss what assumptions should be modified in order to better describe the observed data.

\begin{figure}
    \centering
    \includegraphics[scale=0.4]{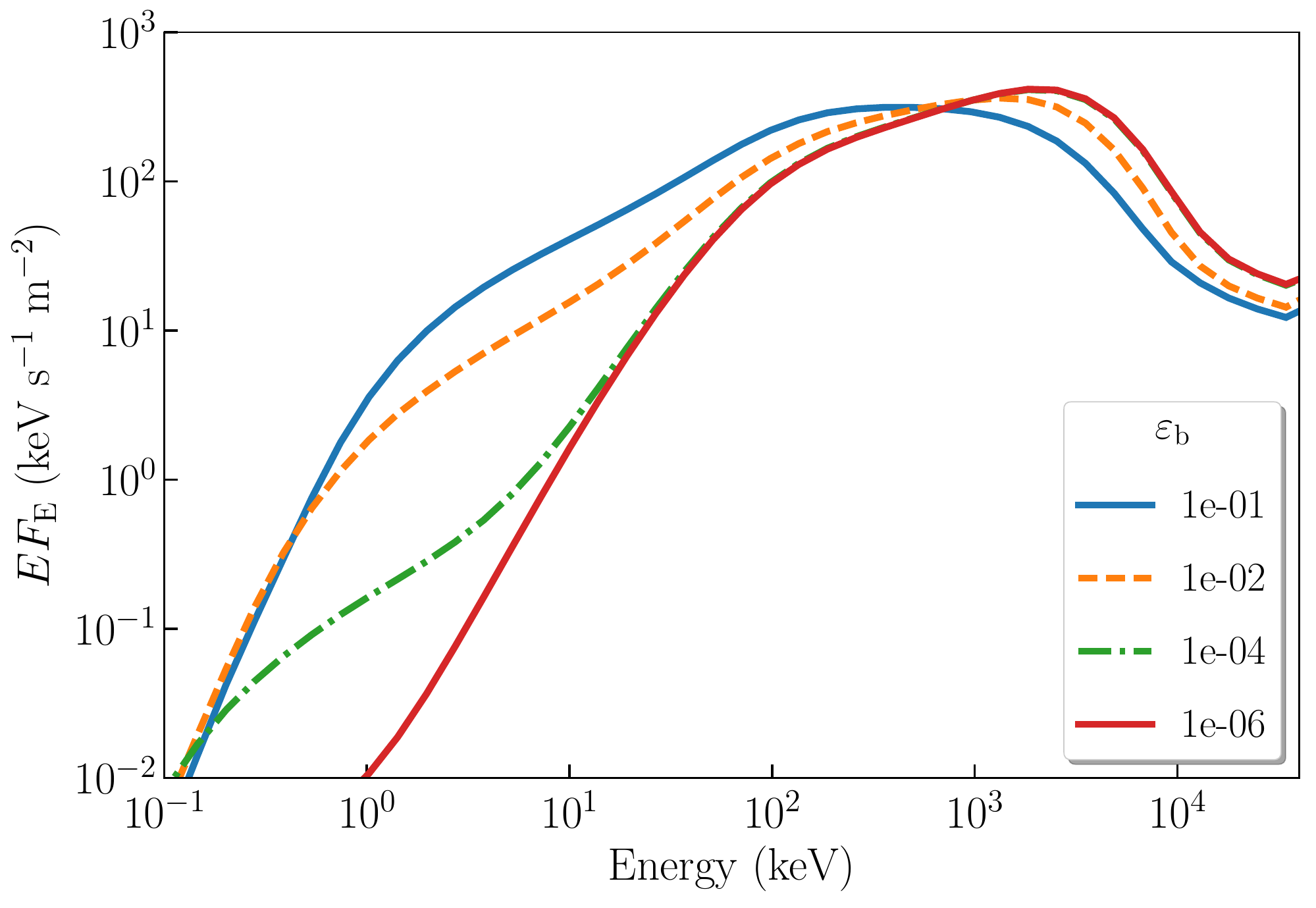}
    \caption{Example of how the model spectra change as a function of the energy dissipated into magnetic fields, $\eb$. The model spectra in this figure were produced with $\ed = 0.1$, $\Lum = 50$ and $\Gamma = 300$. The values of $\eb$ for each spectrum is given by the legend.}
    \label{fig:discussion:synchrotronPhotons}
\end{figure}

Because the parameter space was designed partially based on fits to GBM data, the introduction of XRT data provides new information on what the appropriate parameter space is. Examining the example in Fig.~\ref{fig:res:compareFits_regview}, we note that the XRT data in the left plot are highly reminiscent of what we expect for synchrotron photons in the XRT energy range for our model. In Fig.~\ref{fig:discussion:synchrotronPhotons}, we show examples of what model spectra look like using different values of the magnetization parameter, $\eb$. Although the spectrum in Fig.~\ref{fig:res:compareFits_regview} can be successfully described by our model without synchrotron, the corresponding parameter estimates change. Using a model with synchrotron photons in the XRT energy range might provide an acceptable fit with smaller changes in parameter estimates. This is in no way proof of the presence of synchrotron photons, but it does provide us with strong motivation to expand the parameter space to test a scenario with a significant contribution from synchrotron radiation. 

Additionally, the effects of geometric broadening are expected to be largest in the XRT energy range, \citep{Lundmanetal_2013A_MNRAS}. The general effect of geometric broadening is a softening of the low-energy spectral slope, creating a relative excess of photons at these energies. This again resembles what we see in Fig.~\ref{fig:res:compareFits_regview}, although the effect is expected to be smaller than what we see for $\eb$. Expanding the parameter space in other free parameters, \eg ~the optical depth, $\tau$, could also help improve the fits. Tests conducted in A19 showed that this parameter has only a small impact on the fits. 
However, given the significantly tighter constraints provided by the XRT data, it is possible that $\tau$ would make a difference, since it governs the degree by which the seed blackbody is Comptonized. Particularly, a lower $\tau$ may help describe the double-peaked spectra in GRB~151027A, as discussed above. Thus, although the XRT data suggest that we should reject our current model for several bursts, there is additional parameter space which should be explored before rejecting the physical scenario altogether for these GRBs. The very large number of simulations required for this will be presented in a future work.

As noted above, almost all light curves exhibit some degree of lag between the GBM and XRT data. Apart from being caused by spectral evolution of the prompt emission or by separate emission processes this discrepancy could originate from high latitude emission (see \eg ~\citealt{2007ApJ...666.1002Z}). 
High-latitude emission has a lower Doppler boost than the on-axis emission, which in the framework of our model would be identified as a lower Lorentz factor. Thus, the fact that we consistently find lower values of $\Gamma$ when we include the XRT data is intriguing. 
If there is indeed significant high-latitude emission present in the overlapping time internal of XRT and GBM data, this could help explain the systematic shift to lower $\Gamma$. 
If this is the case it indicates that we cannot neglect the geometry of the emission region at late times, when high latitude emission becomes increasingly prevalent. 

Finally, as discussed in A19, the assumption of internal shocks by setting $\rd = \Gamma^2 r_0$ should be re-evaluated. Letting $r_0$ be a free model parameter would allow us to set the blackbody temperature more freely. 
Since internal shocks are expected to have an efficiency on the order of $1$ - $10$\% \citep{1995Ap&SS.231..441M,1997ApJ...490...92K,1999ApJ...522L.105P} the increased values of $\ed$ when introducing the XRT data are also problematic.
The internal shock assumption is also identified in A19 as a likely cause for the luminosity problem observed there. Thus, in order to further assess the scenario of subphotoshperic dissipation in GRBs, we must drop the assumption that $\rd = \Gamma^2 r_0$.

\section{Summary and conclusions}\label{sec:summary}
We have performed a time-resolved Bayesian analysis of all GRBs that have known redshifts and a significant overlap of GBM and XRT data. This gives us an energy range of $0.3$~keV - $40$~MeV, which encapsulates most distinct spectral features predicted by many physical models. Following the work in A19, we have constructed and tested DREAM1.3, a table model for localized subphotospheric dissipation by internal shocks in a low-magnetization outflow. Our sample consist of 8 GRBs which all have at least one spectrum with significant signal in both the XRT and GBM (SNR$>3$). Binning the data with Bayesian blocks and performing the SNR cut, we obtained 32 spectra. We use a PPC complemented by a manual inspection to assess the quality of the fits. Our main results can be summarized as follows:
\begin{itemize}
    \item 16 of 32 analyzed spectra are well described by the model. No GRB is fully described by the model, but in three GRBs at least half of the fits are accepted. 
    \item The main problem for the model is a small excess of photons at low energies, relative to the extrapolation of the GBM only fits. Even in the case of accepted fits, this leads to systematic changes in parameter estimates when we add the XRT data. Specifically, $\Gamma$ and $\Lum$ decrease whereas $\ed$ increases, albeit within the $3\sigma$ uncertainties of the GBM only fits.
    \item For GRB~151027A the sole rejected bin can likely be explained by the presence of an additional emission component, as previously reported by \cite{2017A&A...598A..23N} and \cite{2018MNRAS.474.2401V}.
    \item The inclusion of XRT data have a large impact on model assessment and leads to much tighter parameter constraints than when using only GBM data. 
\end{itemize}

From these results we see that subphotospheric dissipation with internal shocks cannot describe all prompt emission. However, even for the GRBs where all analyzed spectra are rejected we cannot completely rule out this dissipation scenario. The model only has three free parameters and is thus a limited implementation of the physical scenario. It is encouraging that this simple implementation can describe half of the analyzed spectra. The following are the most promising future improvements:
\begin{itemize}
    \item The introduction of synchrotron photons. This is realized by letting $\eb$ be a free parameter.
    \item Dissipation at different optical depths, characterized by the model parameter $\tau$. In particular, low values of $\tau$ can result in double-peaked spectra (see A15). This is especially relevant for the case of GRB~151027A, where dissipation just below the photosphere is a possible alternative explanation for the rejected spectrum.
    \item The removal of the  assumption of internal shocks ($\rd = \Gamma^2 r_0$), such that $r_0$ is instead set independently of other model parameters. This would decouple the initial blackbody temperature, which sets the position of the low-energy cutoff, from $\Gamma$ and $\Lum$.
\end{itemize}
The work of implementing these improvements will be presented in future work.

\acknowledgments
This work was supported by the G{\"o}ran Gustafsson Stiftelse, the Swedish National Space Board and the Knut \& Alice Wallenberg Foundation. The simulations were performed on resources provided by the Swedish National Infrastructure for Computing (SNIC) at the National Supercomputer Centre (NSC).
FR is supported by the G{\"o}ran Gustafsson Foundation for Research in Natural Sciences and Medicine. This work made use of data supplied by the UK Swift Science Data Centre at the University of Leicester.

%

\vspace{5mm}
\facilities{\textit{Fermi}(GBM), \textit{Swift}(XRT)}


\software{astropy \citep{2013A&A...558A..33A}, PyMultiNest \citep{2014A&A...564A.125B}, Xspec \citep{1999ascl.soft10005A}, Scipy \citep{scipy}, Matplotlib \citep{Hunter:2007}, Pandas \citep{mckinney-proc-scipy-2010}, Seaborn \citep{seaborn}}

\bibliography{bibliography_ahlgren2019b.bib}



\end{document}